\documentclass{PoS}
\usepackage{makerobust}
\MakeRobustCommand\ref
\MakeRobustCommand\cite

\newcommand\avr[1]{\left\langle{#1}\right\rangle}

\title{Fluctuations at finite temperature and density}

\ShortTitle{Fluctuations at finite temperature and density}

\author{\speaker{Szabolcs Bors\'anyi}\thanks{Wuppertal-Budapest collaboration}\\
        Theoretical Physics, Bergsiche Universit\"at Wuppertal, Gaussstr 20, 42119 Wuppertal, Germany\\
        E-mail: \email{borsanyi@uni-wuppertal.de}}

\abstract{
Fluctuations of conserved charges in a grand canonical ensemble can
be calculated as derivatives of the free energy with respect to the
respective chemical potential. They are directly related to experimentally
available observables that describe the hadronization in heavy
ion collisions. The same derivatives can be used to extrapolate zero density
results to finite chemical potential. We review the recent lattice
calculations in the staggered formalism and discuss its implications to
phenomenology and resummed perturbation theory.
}

\FullConference{The 33rd International Symposium on Lattice Field Theory\\
		14 -18 July 2015\\
		Kobe International Conference Center, Kobe, Japan*}

\begin{document}
%]]] -- header

\section{Introduction} %[[[

In Quantum Chromodynamics (QCD) the net number of quarks in a closed system is
conserved flavor by flavor. These conserved charges fluctuate in a grand
canonical ensemble at finite temperature. The magnitude of these fluctuations
are distinctly different in the hadronic and quark gluon plasma phases
\cite{Jeon:2000wg,Asakawa:2000wh}. In heavy ion experiments fluctuations
appear in the event-by-event statistics of the net charges \cite{Luo:2015cea}.
Continuum extrapolated lattice simulations with physical quark masses, that
have already determined the QCD transition temperature
\cite{Aoki:2006br,Aoki:2009sc,Borsanyi:2010bp,Bazavov:2011nk} and the equation
of state \cite{Borsanyi:2010cj,Borsanyi:2013bia,Bazavov:2014pvz}, are now 
used to calculate these fluctuations in the grand canonical ensemble.
The actual comparison of STAR data
to lattice have placed the chemical freeze-out (i.e. the instant of the last
inelastic scattering) at or slightly below the transition temperature
\cite{Borsanyi:2014ewa}.  Higher order baryon fluctuations are also indicators
for the closeness of a critical end point \cite{Stephanov:1999zu}.

From the theoretical point of view fluctuations are a proxy for the comparison
between various theoretical and model approaches, such as the Hard Thermal Loop
(HTL) perturbation theory at high temperature
\cite{Braaten:1991gm,Blaizot:2001vr,Andersen:2002ey}, the Hadron Resonance
Gas (HRG) model in the confined phase \cite{Dashen:1969ep,Venugopalan:1992hy},
or improved low energy models in the transition region \cite{Fu:2015naa}.

Fluctuations are formally equivalent to the Taylor expansion coefficients
of the free energy in a grand canonical ensemble. Thus, finite density methods
open new ways to calculate these \cite{Gattringer:2014hra,Fukuda:2015mva}. On
the other hand, the extrapolation of, for example, the equation of state to
finite chemical potential is often a by-product of the calculation of generic
fluctuations \cite{Borsanyi:2012cr,Hegde:2014wga}. The acute interest in
small-$\mu$ physics was highlighted at this conference by four lattice
groups presenting their results on the curvature of the chiral transition
line in the QCD phase diagram \cite{Bonati:2015bha,Bellwied:2015rza,Cea:2015cya,Hegde:2015tbn}. These results have been confronted to heavy ion data,
where the experimentally observed fluctuations were matched to lattice data
\cite{Bazavov:2015zja}. This comparison has shown a slight tension: the 
fluctuation data prefer a smaller (or even negative) curvature, which was
also observed in a HRG-based calculation earlier \cite{Alba:2015lxa}.
The phase diagram at small chemical potential is likely to stay a hot topic
at the coming Lattice and Quark Matter conferences.

In this work I review the recent progress of the fluctuation program on the
lattice. Few years ago finite temperature results for the second order
fluctuations became available in the continuum limit with physical quark masses
\cite{Borsanyi:2011sw,Bazavov:2012jq}.  Since then, several
higher order fluctuations have been continuum extrapolated, too,
\cite{Borsanyi:2012rr,Borsanyi:2013hza,Bellwied:2013cta,Bazavov:2013uja,Bellwied:2015lba,Ding:2015fca}.
Sophisticated combinations of higher order cross-correlators of conserved
charges were also used to constrain the QCD spectrum in the strange
\cite{Bazavov:2014xya} and in the charm sectors \cite{Bazavov:2014yba}, as well
as to study the pattern of melting of hadronic bound states \cite{Bazavov:2013dta,Mukherjee:2015mxc}.
The papers cited in this paragraph were using staggered quarks. It must be
said, that, although not yet with physical quark masses, second order
fluctuations have already been calculated with Wilson quarks as well
\cite{Borsanyi:2012uq,Borsanyi:2015waa,
Giudice:2013fza,Aarts:2014nba,Gattringer:2014hra}, a continuum limit
was possible even in the overlap formulation \cite{Borsanyi:2015zva}.

I structured this presentation around five pillars:
I am starting with the technical difficulties lattice groups face before a
continuum extrapolated result can be obtained. Then I am presenting comparisons
with the Hadron Resonance Gas model, and, in the following section, with
improved perturbation theory. Then I am discussing the relations to non-zero
density physics. Finally, since fluctuations are experimentally measured
quantities a comparison between theory and experiment is presented.

%]]]

\section{Fluctuations on the lattice\label{sec:lat}} %[[[

The dimensionless fluctuations and correlators
of quark numbers are the derivatives of the free energy ($\sim$ pressure)
with respect to the respective chemical potentials:
\begin{equation}
\chi^{u,d,s,c}_{i,j,k,l}= \frac{\partial^{i+j+k+l} (p/T^4)}{
(\partial \hat\mu_u)^i
(\partial \hat\mu_d)^j
(\partial \hat\mu_s)^k
(\partial \hat\mu_c)^l
}
\label{eq:pderiv}
\end{equation}
with $\hat\mu_q=\mu_q/T$. These are also called (generalized)
susceptibilities.

\begin{figure}[b]
\begin{center}
\includegraphics[height=2in]{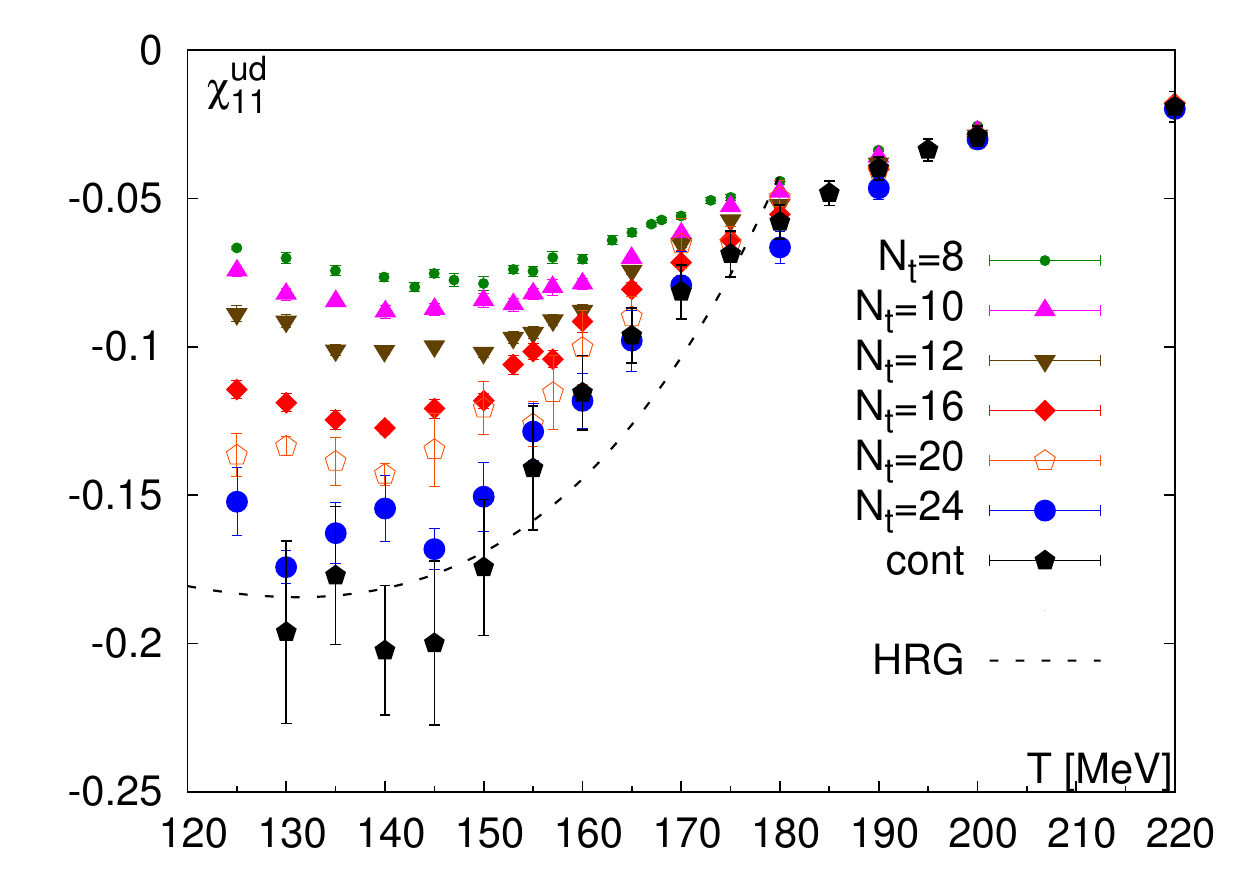}
\includegraphics[height=2in]{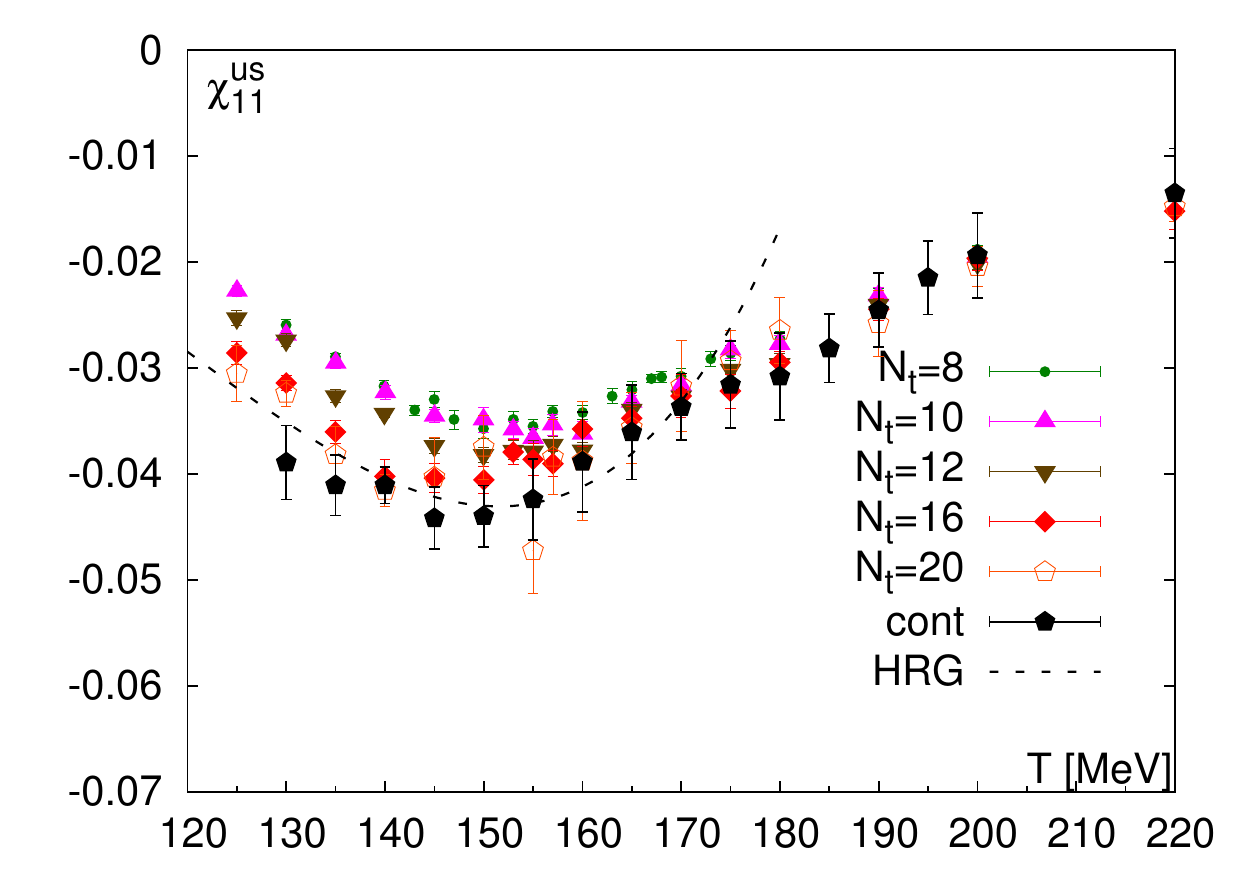}
\end{center}
\caption{\label{fig:c2ud}
The up-down correlator ($\chi^{ud}_{11}$) for a range of lattice spacings 
using the 4stout staggered action of the Wuppertal-Budapest group
\cite{Bellwied:2015lba}.
We also show the light-strange correlator ($\chi^{us}_{11}$) in the right panel, for comparison.
}
\end{figure}

$\chi^{ud}_{11}$, for example, expresses the number
of up quarks generated by an infinitesimal change in the down
quark chemical potential $\hat\mu_d$ in a volume of $T^{-3}$. 
For this observable we present the lattice data of the Wuppertal-Budapest
collaboration and the corresponding continuum limit in the left panel of 
Fig.~\ref{fig:c2ud}, a continuum result that came 14 years after its first
lattice computation \cite{Gavai:2001ie}.
Clearly, if light mesons (pions) dominate the thermodynamics at low
temperature, one expects $\chi^{ud}_{11}$ to be zero at $T=0$ and negative at
$T\lesssim T_c$. As the $T/M_{\rm proton}$ ratio becomes non-negligible with
increasing temperature light baryons (nucleons) will coexist with pions and the
baryonic contribution will reverse the trend in $\chi^{ud}_{11}(T)$. The
decoupling of quarks flavors is one manifestation of deconfinement. In
perturbation theory the correlation drops to zero as $\alpha^3\log\alpha$.
We will discuss the comparison to the HRG model in section~\ref{sec:hrg}
and study the high temperature behaviour in section~\ref{sec:htl}.

Chemical potentials are introduced as imaginary, homogeneous, abelian $A_0$
fields on the lattice. Only the fermion determinant of the respective flavor
depends on it. Thus, the observable $D_j$ to be calculated is the $j$-th
derivative of the logarithm of the fermion determinant ($M$), e.g.
$D_1=\frac{1}{4}\mathrm{Tr} [ M^{-1} dM/d\mu]$. (The factor $1/4$ is only
present for staggered fermions.) It is now straightforward to write down the
lattice observable for the fluctuations. First we remark, that
$\chi^u_1(T) \sim d\log Z/d\mu_u \sim \avr{D_1^u}$ is zero at vanishing chemical potential due to the $\mathcal{C}$-symmetry, and the same holds for all
odd derivatives.
\begin{eqnarray}
\chi^{u}_2&=&\frac{1}{TV} \left [ \avr{D_1^uD_1^u} - \avr{D_1^u}\avr{D_1^u}
+\avr{D_2^u} \right]
,\label{eq:chi2u}\\
\chi^{ud}_{11}&=&\frac{1}{TV} \left [ \avr{D_1^uD_1^d} - \avr{D_1^u}\avr{D_1^d} 
\right]\,.\label{eq:chi2ud}
\end{eqnarray}
In other words, if we assume degenerate up and down quarks
(and no dynamical QED effects), then $\chi_{11}^{ud}$ will be the disconnected
part of the light quark number susceptibilities. The traces in the $D_j$
contributions are typically calculated using Gaussian random sources, excluding,
of course, those contributions from $\avr{D_1^uD_1^u}$ where both traces were
estimated using the same random source.
For details, see Refs.~\cite{Gavai:2001ie,Allton:2002zi} or 
the more recent \cite{Bellwied:2015lba,Ding:2015fca}.

Notice the strong lattice spacing dependence in Fig.~\ref{fig:c2ud}.
The Wuppertal-Budapest group has gone as far as including $N_t=24$
lattices into the continuum extrapolation. At the heart of the
discretization errors there is the typically slow approach of the
staggered pions to the physical spectrum. Observables, like
$\chi^{u}_2+\chi^{ud}_{11}$, on the other hand, are related to light
baryon fluctuations and have a more favourable continuum scaling.

It was pointed out in Ref.~\cite{Allton:2002zi} that the odd $D_j$ coefficients
are the Taylor coefficients of the phase of the fermion determinant $\mathrm{det}~M=|\mathrm{det}~M|e^{i\theta}$:
\begin{equation}
\theta = N_f\mathrm{Im}~\left[
\frac{\mu_B}{3}  D_1
+ \frac{\mu_B^3}{3^3\cdot 3!}  D_3
+\dots\right]\,
\end{equation}
where $N_f$ are the number of flavors for which the chemical potential
is introduced. With the choice of $N_f=2$ light quarks we have
\begin{equation}
\avr{e^{i\theta}} = 1+ \frac{2}{9}\mu_B^2 V T \chi^{ud}_{11} +\mathcal{O}(\mu_B^4)\,.
\end{equation}
This clearly shows, that the severity of the sign problem for small chemical
potentials is basically set by $\chi^{ud}_{11}$. Notice that staggered lattice
artefacts, actually, greatly ease the sign problem. Closer to the continuum
limit, $\chi^{ud}_{11}\approx -0.2$, to leading order $\avr{e^{i\theta}}=0.5$
is reached at $\mu_B=184$, $100$, $65$~MeV for an aspect ratio of $LT=2$, 3 and
4, respectively.

\begin{figure}[t]
\begin{center}
\includegraphics[width=2.9in]{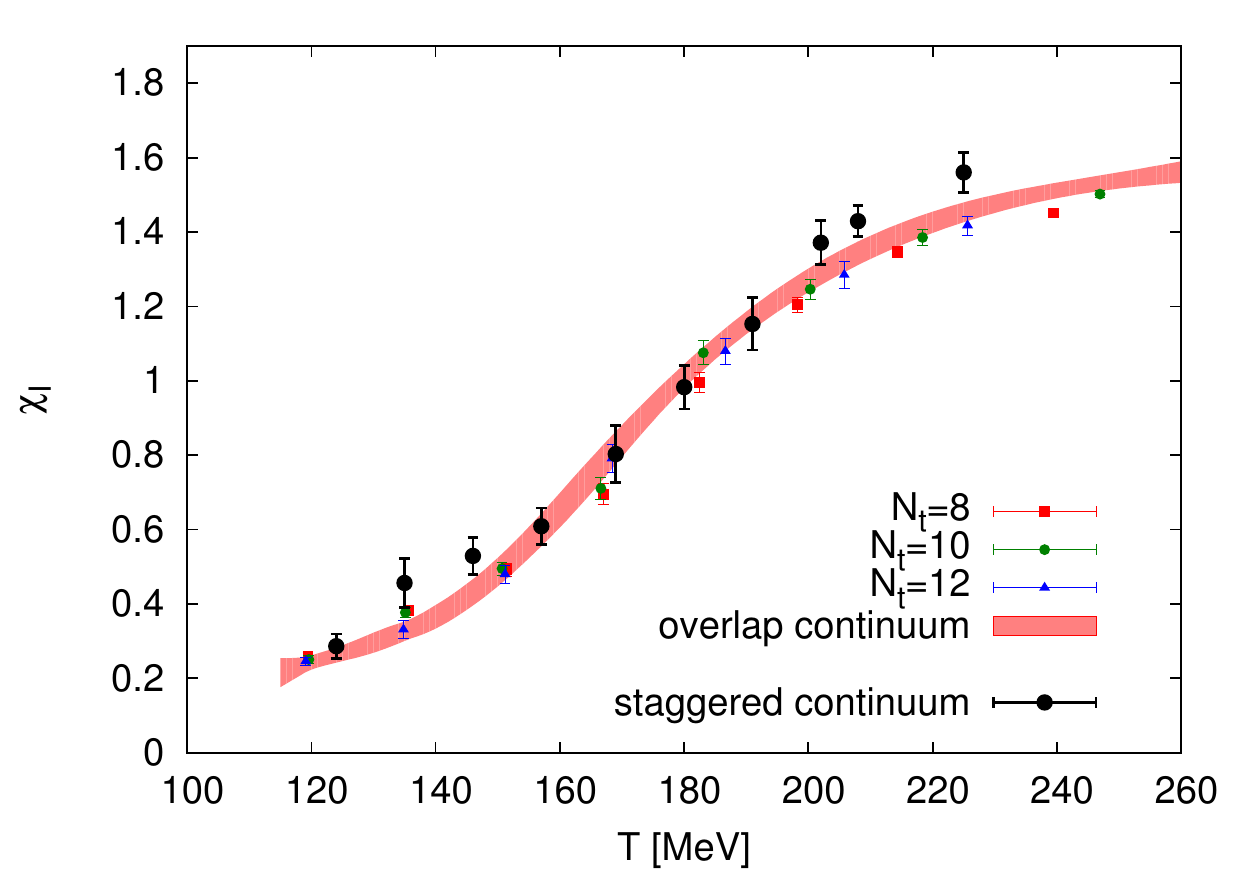}
\end{center}
\caption{\label{fig:ovc2I}
Continuum extrapolation of the isospin fluctuations in an $N_f=2$ theory
with $M_\pi=350~\mathrm{MeV}$ using chiral fermions. The staggered
result is also continuum extrapolated and corresponds to the same theory
\cite{Borsanyi:2015zva}.
}
\end{figure}

The connected contribution $\avr{D^{u}_2}$ in Eq.~\ref{eq:chi2u} is
proportional to the isospin fluctuations. Being dominated by pions
at low temperature, this quantity shows the same (absolute) discretization
errors as $\chi^{ud}_{11}$. This difficulty, however, can be overcome
by switching to a chiral representation of fermions. This has been done
in Ref.~\cite{Borsanyi:2015zva}, showing results with two dynamical
flavors with $M_\pi=350$~MeV, (see Fig.~\ref{fig:ovc2I}).

\begin{figure}[t]
\begin{center}
\includegraphics[height=2in]{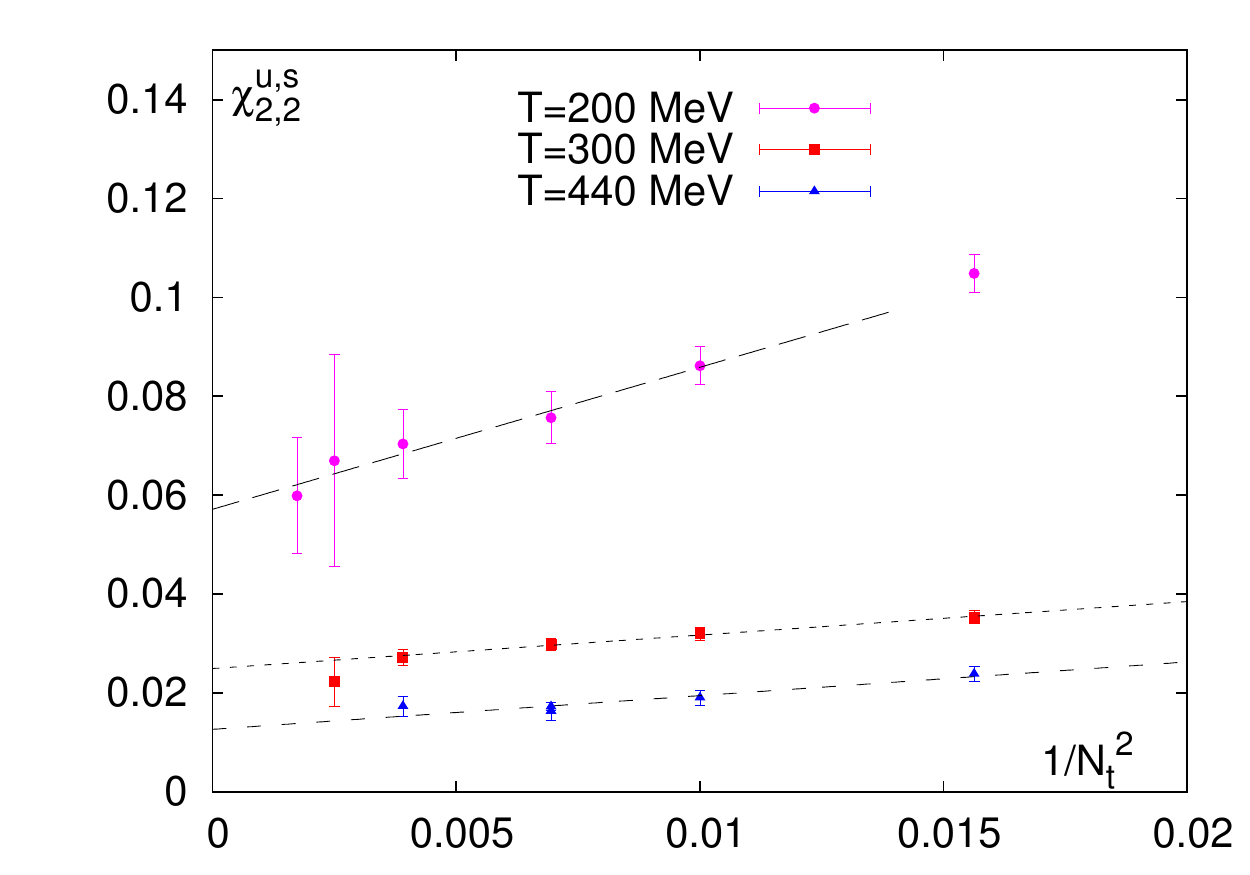}
\includegraphics[height=2in]{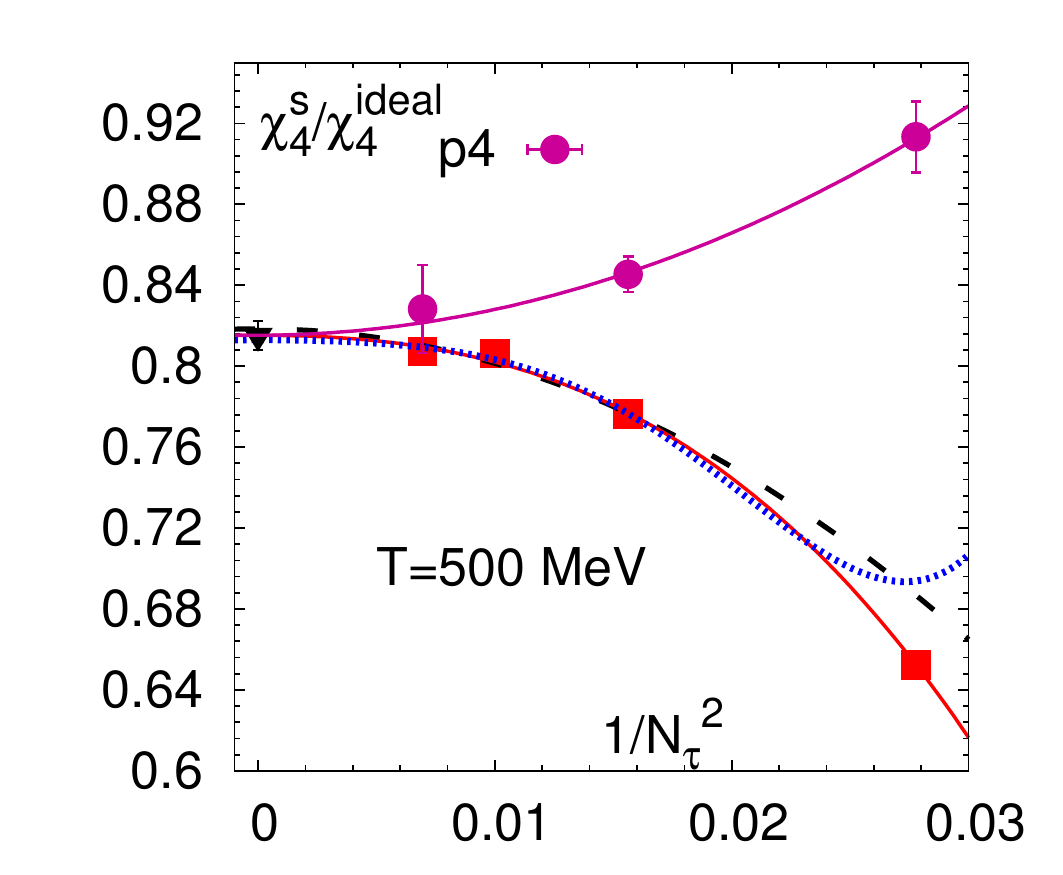}
\end{center}
\caption{\label{fig:hotextra}
Continuum scaling of the off-diagonal and diagonal higher order fluctuations.
On the left panel we show $\chi^{us}_{22}$ for the 4stout staggered action
of the Wuppertal-Budapest group \cite{Borsanyi:2015zva}, the right
panel shows the scaling of the HISQ and p4 actions for $\chi^{s}_4$ by
the BNL group \cite{Ding:2015fca}.
}
\end{figure}

The continuum scaling for higher order fluctuations are under control
at high temperatures (see Fig.~\ref{fig:hotextra}). Around and below
the transition temperature lattice calculations face two challenges:
the fourth moment of the electric charge fluctuation has strong lattice
artefacts with all known staggered actions, thus finer lattices are needed.
On the other hand, higher order baryon fluctuations are very noisy at low $T$,
especially on fine lattices, where disconnected contributions are
not suppressed by lattice artefacts (see Fig.~\ref{fig:c2ud}).
Disconnected parts come with less noise in smaller volumes. However,
finite volume effects can be present in the continuum, even though
they are often hidden at finite $N_t$ because of the same lattice artefacts.

%]]] sec:lat

\section{Fluctuations and the Hadron Resonance Gas model\label{sec:hrg}} %[[[

At low temperatures QCD can be modelled as a gas of uncorrelated hadrons,
where interactions are included as a tower of resonances. These states
are taken from the Particle Data Book \cite{Agashe:2014kda}. The pressure
of the Hadron Resonance Gas reads \begin{equation}
\frac{p^\mathrm{HRG}}{T^4}=\frac{1}{VT^3}
+\sum_{i\in\mathrm{mesons}} \log \mathcal{Z}^M(T,V,m_i,\{\mu\})
+\sum_{i\in\mathrm{baryons}} \log \mathcal{Z}^B(T,V,m_i,\{\mu\})
\end{equation}
with
\begin{eqnarray}
\log \mathcal{Z}^{M/B}_{m_i} &=&
\mp \frac{V d_i}{2\pi^2}\int_0^\infty dk k^2 \log\left(1\mp z_i e^{-\sqrt{m_i^2+k^2}/T}\right)\\
&=&\frac{VT^3}{2\pi^2}d_i \frac{m_i^2}{T^2} \sum_{k=1}^\infty
(\pm)^{k+1} \frac{z_i^k}{k^2} K_2(k m_i/T)\,,
\label{eq:hrgZ}
\end{eqnarray}
with the fugacity factor {\small
$z_i=\exp(B_i\hat\mu_B+Q_i\hat\mu_Q+S_i\hat\mu_S)$}, the degeneracy factor
$d_i$ and {\small $\hat\mu=\mu/T$}.

In HRG as well as in phenomenology it is more convenient to work in the $B-Q-S$
base, already used in Eq.~(\ref{eq:hrgZ}). These refer to the baryon number,
electric charge and strangeness, the corresponding chemical potentials are
defined by the following equations:
\begin{equation}
\mu_u=\frac13\mu_B+\frac23\mu_Q\,,\quad
\mu_d=\frac13\mu_B-\frac13\mu_Q\,,\quad
\mu_s=\frac13\mu_B-\frac13\mu_Q-\mu_S\,.
\end{equation}
The six second order derivatives in the $B-Q-S$ base are mapped into
the $u-d-s$ base where due to the $u\leftrightarrow d$ degeneracy of the lattice
setup only four derivatives are different:
\begin{eqnarray}
\chi_2^{u}&=&2\chi_2^{B}+\chi_2^{Q}+\chi_{11}^{BS}\,,\\
\chi_2^{s}&=&\chi_2^{S}\,,\\
\chi_{11}^{ud}&=&\frac{5}{2}\chi_2^{B}-\chi_2^{Q}+\frac{1}{2}\chi_2^{S}+2\chi_{11}^{BS}\,,\\
\chi_{11}^{us}&=-&\frac{1}{2}\chi_2^{S}-\frac{3}{2}\chi_{11}^{BS}
=-3\chi_{11}^{QS}+\chi_2^S
=\frac{3}{2}\chi_2^{B}-\frac{1}{2}\chi_2^{S}-3\chi_{11}^{BQ}\,,
\end{eqnarray}

\begin{figure}[t]
\begin{center}
\includegraphics[height=2in,bb=90 70 750 530]{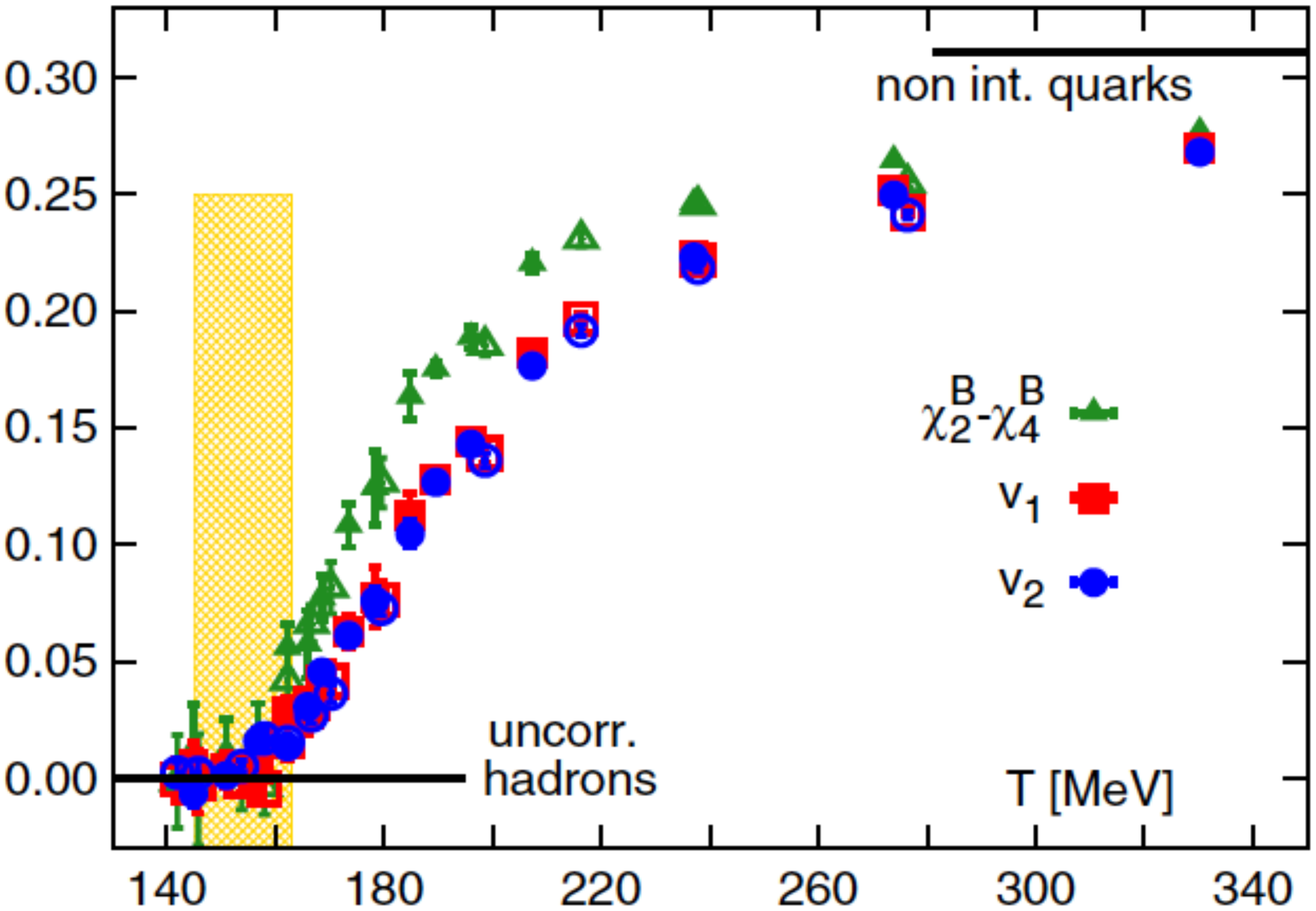}
\includegraphics[height=1.96 in]{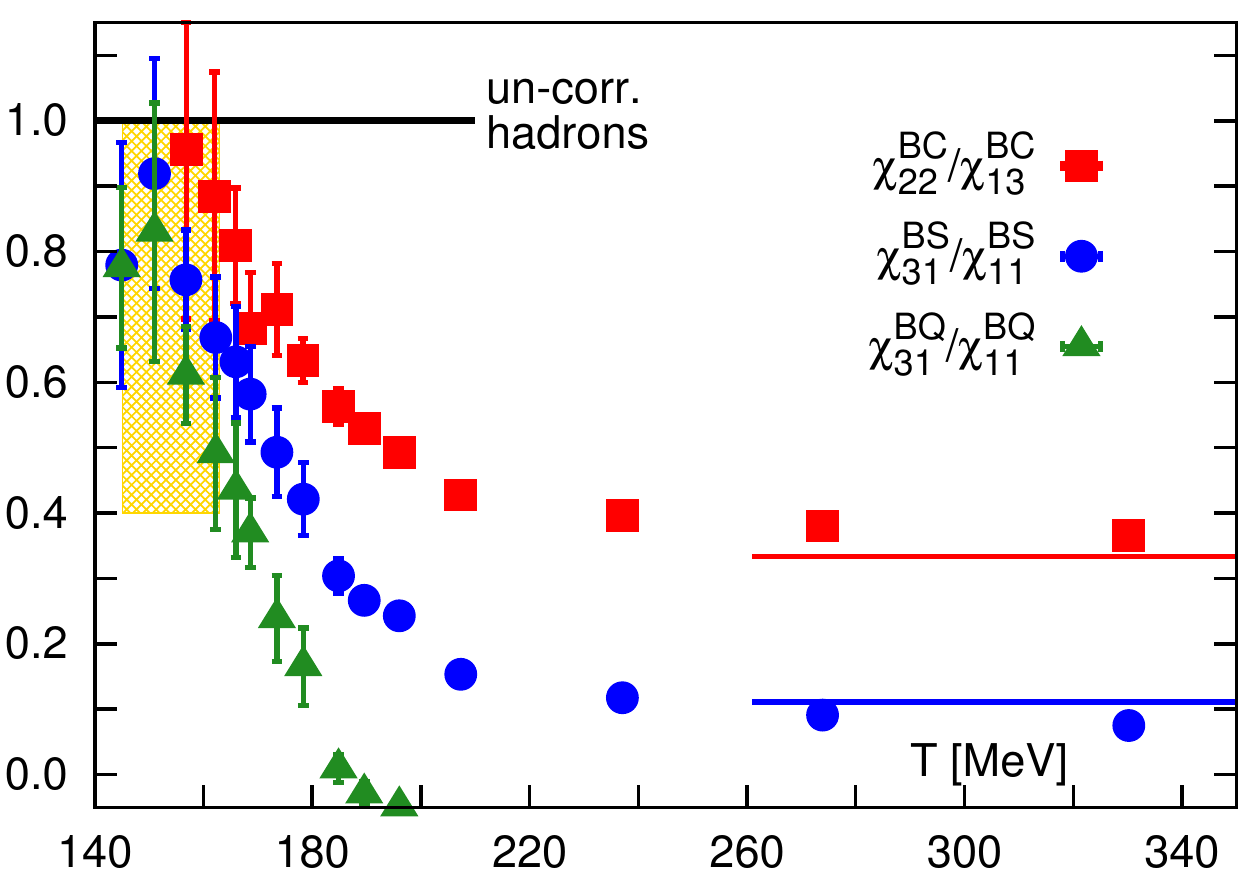}
\end{center}
\caption{\label{fig:hrgidentities}
Combinations of generalized susceptibilities that are constant 0 (left panel)
or 1 (right panel) within the Hadron Resonance Gas model for any list of
resonances. Calculating these in full QCD the departure of these curves from
the hadronic expectation provides for a flavor dependent indicator of
deconfinement. On the right hand side the horizontal lines show the Stefan-Boltzmann limit according to Eq.~(\ref{eq:freequarks}).
Here we show the plots by the BNL-Bielefeld group
\cite{Bazavov:2013dta,Bazavov:2014yba}.  }
\end{figure}

The first comparison between lattice and HRG has already been shown in
Fig.~\ref{fig:c2ud}. The two panels show an agreement both for the light-light
and the light-strange correlators, yet the highest temperature where
the hadronic description is consistent with data is slightly higher for the
strange at the level of this precision. The differences between
light and strange sectors have been put to a less ambiguous test.
Certain combinations of derivatives are constant zero (or $=1$) in HRG,
independently of the resonance list in use, but are non-zero (or $\ne1$) if
quarks are free. These typically involve the difference (or ratio) of
correlators with different order of baryon derivative. For baryons, typically
only the $k=1$ term contributes significantly in Eq.~(\ref{eq:hrgZ}),
neglecting
$k\ne1$ is called the Boltzmann approximation. In this limit the
contribution of each resonance of mass $m_i$ and its antiparticle
to the QCD pressure is
\begin{equation}
\frac{p}{T^4} = 2d_i \frac{VT^3}{2\pi^2} \frac{m_i^2}{T^2} K_2(m_i/T)
\cosh(B_i \hat\mu_B + Q_i \hat \mu_Q + S_i\hat \mu_S),\
\end{equation}
where $B_i$, $Q_i$ and $S_i$ are integer quantum numbers. For an ideal gas,
on the other hand, the pressure at finite chemical potential reads
\cite{kapusta:book}
\begin{equation}
\frac{p}{T^4}= \frac{8\pi^2}{45} + \frac{7\pi^2}{60} N_f +
\frac12 \sum_f \left( \frac{\mu_f^2}{T^2} + \frac{\mu_f^4}{2\pi^2T^4}\right)\,.
\label{eq:freequarks}
\end{equation}
There are combinations of fluctuations that have a non-zero Stefan-Boltzmann
limit but vanish in the HRG approach:
\begin{eqnarray}
\chi^{B}_4-\chi^B_{2} &=& 0\,,\\
v_1=\chi^{BS}_{31}-\chi^{BS}_{11} &=& 0\,,\\
v_2=\frac{1}{3}(\chi^S_2-\chi^S_4)
-2\chi^{BS}_{13}-4\chi^{BS}_{22}-2\chi^{BS}_{31}&=&0\,.
\end{eqnarray}
Thus, these can be considered as indicators for deconfinement. Others, 
$\chi^{BC}_{11}/\chi^{BC}_{13}$, 
$\chi^{BS}_{31}/\chi^{BS}_{11}$ and
$\chi^{BQ}_{31}/\chi^{BQ}_{11}$ are always $=1$ in the confined phase. We show
the results for these combinations in Fig.~\ref{fig:hrgidentities} using
the $N_t=8$ data of the BNL-Bielefeld group. Notice that both
$\chi_2^B-\chi_4^B$ and $\chi^{BQ}_{31}/\chi^{BQ}_{11}$ are dominated by the
light flavors. The plots in Fig.~\ref{fig:hrgidentities} suggests a slight
flavor dependence in the deconfinement pattern. In Ref.~\cite{Bellwied:2013cta}
the Wuppertal-Budapest group has presented the light counterparts of 
$v_1$ and $v_2$, showing a slight difference between flavors.

%]]] sec:hrg

\section{Fluctuations and improved perturbation theory\label{sec:htl}}%[[[

In the high temperature limit perturbation theory gives an appropriate
description of QCD. The pressure at zero chemical potential is known
up to $\alpha^3\log(\alpha)$ order \cite{Kajantie:2002wa}, which
was extended to $\mu_B>0$ in \cite{Vuorinen:2003fs,Ipp:2006ij}.
Combining these findings with the dimensional reduction \cite{Blaizot:2003iq}
recently estimates for the four-loop perturbative quark number
susceptibilities have emerged \cite{Andersen:2012wr,Mogliacci:2013mca}.

The reorganization of the perturbative series around hard thermal loops
also improves the convergence \cite{Braaten:1991gm,Andersen:2002ey}. This
was exploited in full QCD in Refs.~\cite{Strickland:2010tm,Andersen:2011ug}.
For the fluctuations subsequent orders have been calculated in Refs.~\cite{Blaizot:2001vr,Blaizot:2002xz,Andersen:2012wr,Haque:2013qta,Haque:2013sja}.
These developments allowed the HTL calculation of the equation of state
at finite chemical potentials \cite{Haque:2014rua}.

We expect that at several $T_c$ temperature these diagrammatic approaches
agree among themselves and are consistent with lattice. To find out
the range of validity both the Wuppertal-Budapest and BNL groups
have conducted high temperature simulations and calculated generalized
quark number susceptibilities in the continuum limit
\cite{Borsanyi:2012rr,Bellwied:2015lba,Bazavov:2013uja,Ding:2015fca}.

\begin{figure}[ht]
\begin{center}
\includegraphics[width=2.9in]{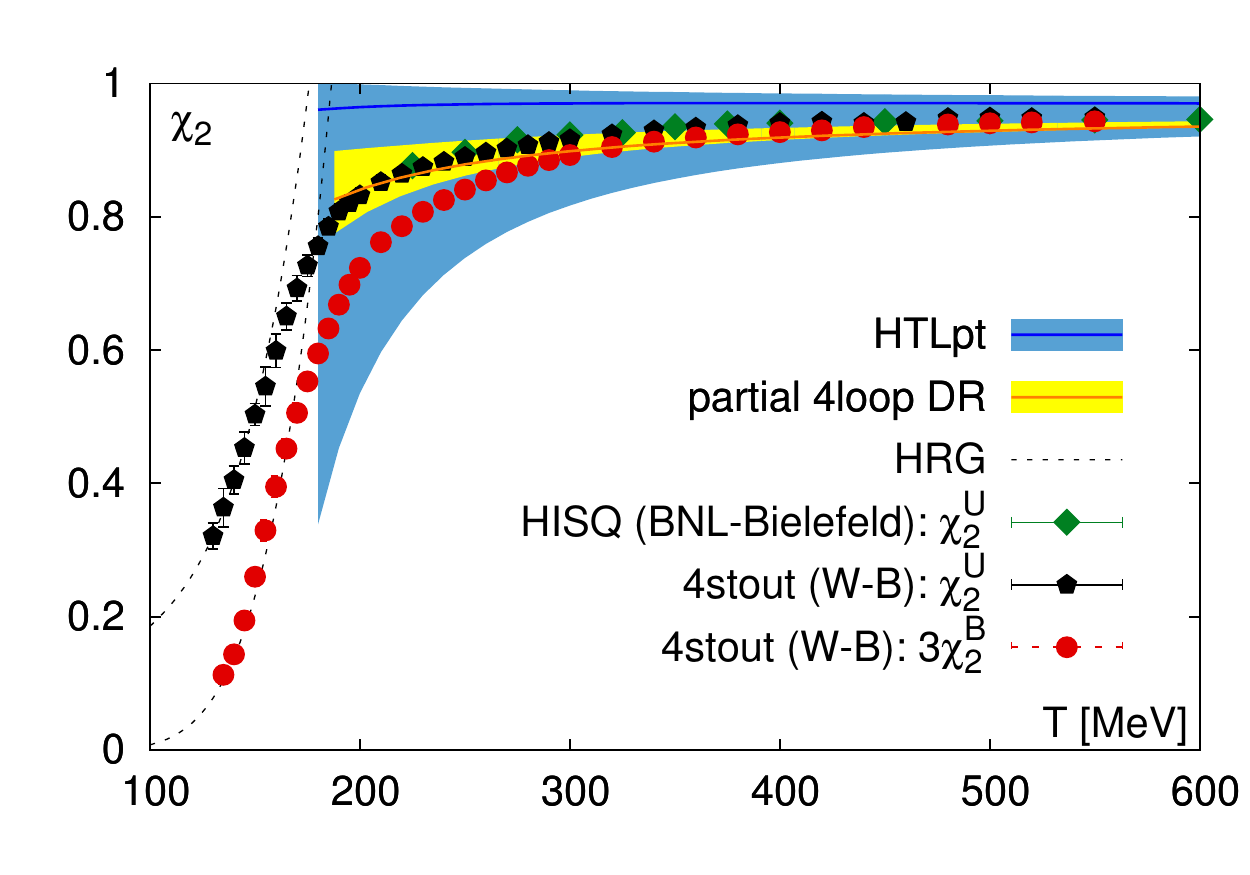}
\includegraphics[width=2.9in]{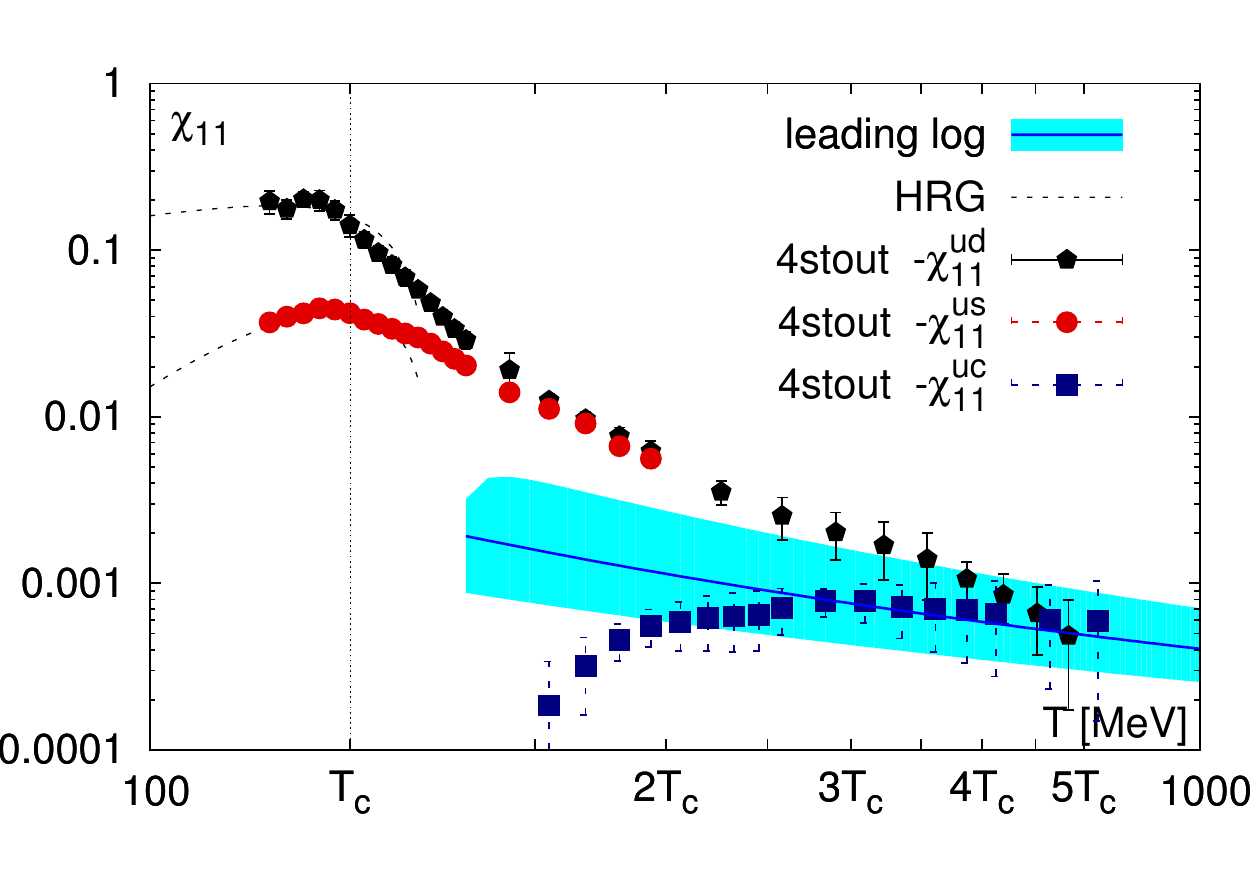}
\end{center}
\caption{\label{fig:c2U}
Left: The (diagonal) light quark number susceptibility $\chi^U_2$ and the
baryon susceptibility $\chi^B_2$ at high temperatures
\cite{Bellwied:2015lba,Bazavov:2013uja}.
We also show the latest (improved) perturbative results with Hard Thermal Loops
(HTL) \cite{Haque:2013sja} and dimensional reduction (DR)
\cite{Mogliacci:2013mca}. 
\label{fig:c2ud_hot}
Right: The (off-diagonal) quark correlators between the light quark and the
light (black), strange (red) and charm (blue) quarks.  The light-light
correlator spans more than two orders of magnitude in the temperature range
between $T_c$ and $5T_c$.  The leading $\mathcal{O}(\alpha^3\log\alpha)$
perturbative result comes from Ref.~\cite{Blaizot:2001vr}.  
}
\end{figure}

In the left panel of Fig.~\ref{fig:c2U} we show $\chi_2$ for the light quark as
well as for the baryons. The Wuppertal-Budapest data is based on the 4stout
action \cite{Bellwied:2015lba}, the BNL-Bielefeld result is using the HISQ
action (in combined analysis with p4 data) \cite{Bazavov:2013uja}.

The off-diagonal susceptibilities (already shown in Fig.~\ref{fig:c2ud}) are
compared to the leading log result in right panel of Fig.~\ref{fig:c2ud_hot}.
The magnitude of the Hadron Resonance Gas result on $\chi_{11}$ is two orders
of magnitude higher the perturbative estimate ($\sim\alpha^3\log(\alpha)$),
which is reached at about $3T_c$ temperature by the lattice data. It is
remarkable that even the charm-up correlator is consistent with the
perturbative result in the temperature range accessible to simulations.  Below
$3T_c$ the charm is practically uncorrelated with the other quarks.  The mass
of the strange quark becomes negligible near $1.5T_c$.

\begin{figure}[ht]
\begin{center}
\includegraphics[width=2.9in]{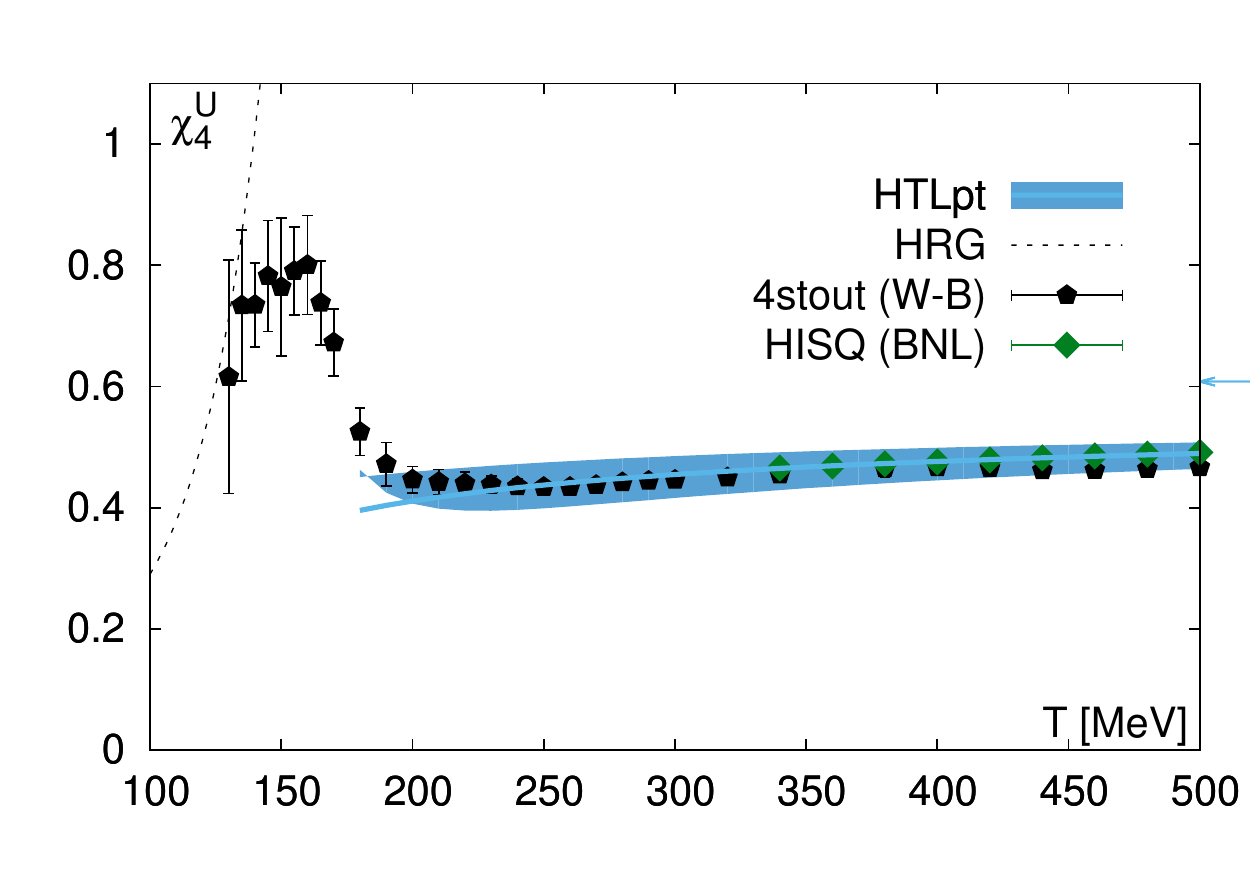}
\includegraphics[width=2.9in]{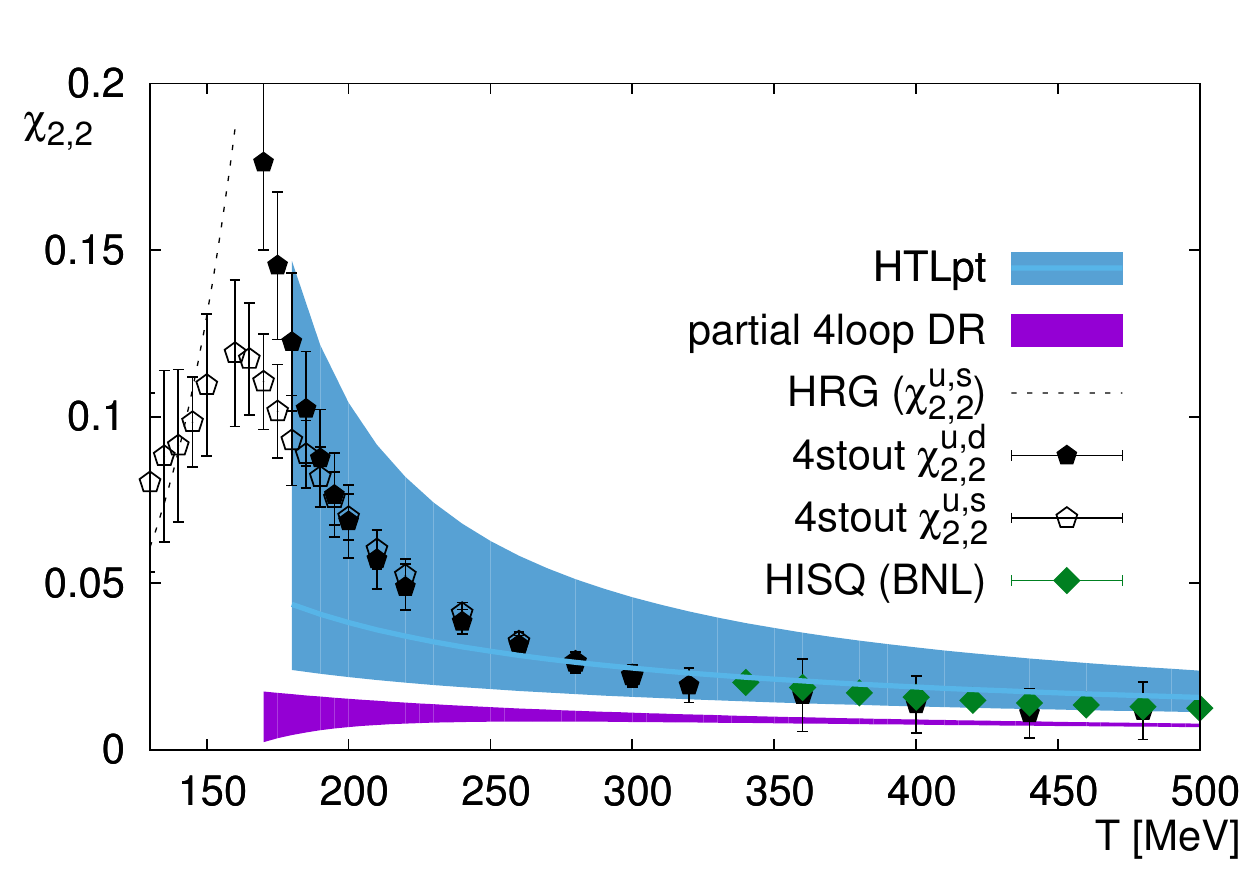}
\end{center}
\caption{\label{fig:chi4}
The diagonal (left) and off-diagonal (right) fourth order fluctuation at high
temperature. We compare the continuum result of two collaborations:
Wuppertal-Budapest \cite{Bellwied:2015lba} and BNL \cite{Ding:2015fca}. The
shown off-diagonal derivative is the only one with a non-vanishing contribution
in three-loop HTL \cite{Haque:2014rua}. The dimensional reduction (DR) data set
is from Ref.~\cite{Mogliacci:2013mca}.
}
\end{figure}

Finally we show the diagonal and off-diagonal fourth order correlators.
In Fig.~\ref{fig:chi4} we show $\chi^U_4$ on the left panel and
both the light-light and the light-strange correlators ($\chi_{22}$) on the right panel.  Here the effect of the strange mass diminishes already
at around 200 MeV temperature. 
The agreement with the HTL result starts at a low temperature because
of the large uncertainties, the central line is approached at $T\sim 250$ MeV.
We also show the prediction of dimensional reduction \cite{Mogliacci:2013mca},
though for $\chi_{22}$ it is not in agreement with HTL, nor with the lattice data.

%]]] sec:htl

\section{Fluctuations and physics at non-zero chemical potential\label{sec:mub}}%[[[

Fluctuations naturally give access to the physics at small chemical potential.
The definition of the generalized susceptibilities ($\chi$) in
Eq.~(\ref{eq:pderiv}) immediately connect the Taylor coefficients of the
QCD pressure to the baryon fluctuations. Continuum results have
been calculated by the Wuppertal-Budapest group as well as the HotQCD
collaboration to leading order \cite{Borsanyi:2011sw,Bazavov:2012jq}, to
next-to-leading order the Wuppertal-Budapest group published
Refs.~\cite{Borsanyi:2013hza,Bellwied:2015lba}.

For the calculation of the equation of state at small chemical potential
for the phenomenological use we still need to add one ingredient.
Since there is no net strangeness in the colliding nuclei, and the
strangeness is conserved, the expectation value of the net strangeness
is zero in experiment. This must be reflected in the grand canonical ensemble
that we are using to calculate the bulk thermodynamics of the plasma.

Thus a pair of equations can be set up and solved, requiring that at any finite
chemical potential
\begin{equation}
\chi^S_1(\mu_B,\mu_Q,\mu_S) = 0\,,\qquad
\chi^B_1(\mu_B,\mu_Q,\mu_S) = \frac{Z}{A}
\chi^Q_1(\mu_B,\mu_Q,\mu_S)\,.\label{eq:neutrality}
\end{equation}
These two equations connect the electric charge and strangeness chemical
potentials to the baryon chemical potential, which have been
calculated to next-to-leading order using fourth-order fluctuations
\cite{Bazavov:2012vg,Borsanyi:2013hza} and for imaginary chemical
potentials \cite{Bellwied:2015rza}.

\begin{figure}[ht]
\begin{center}
\includegraphics[width=2.9in]{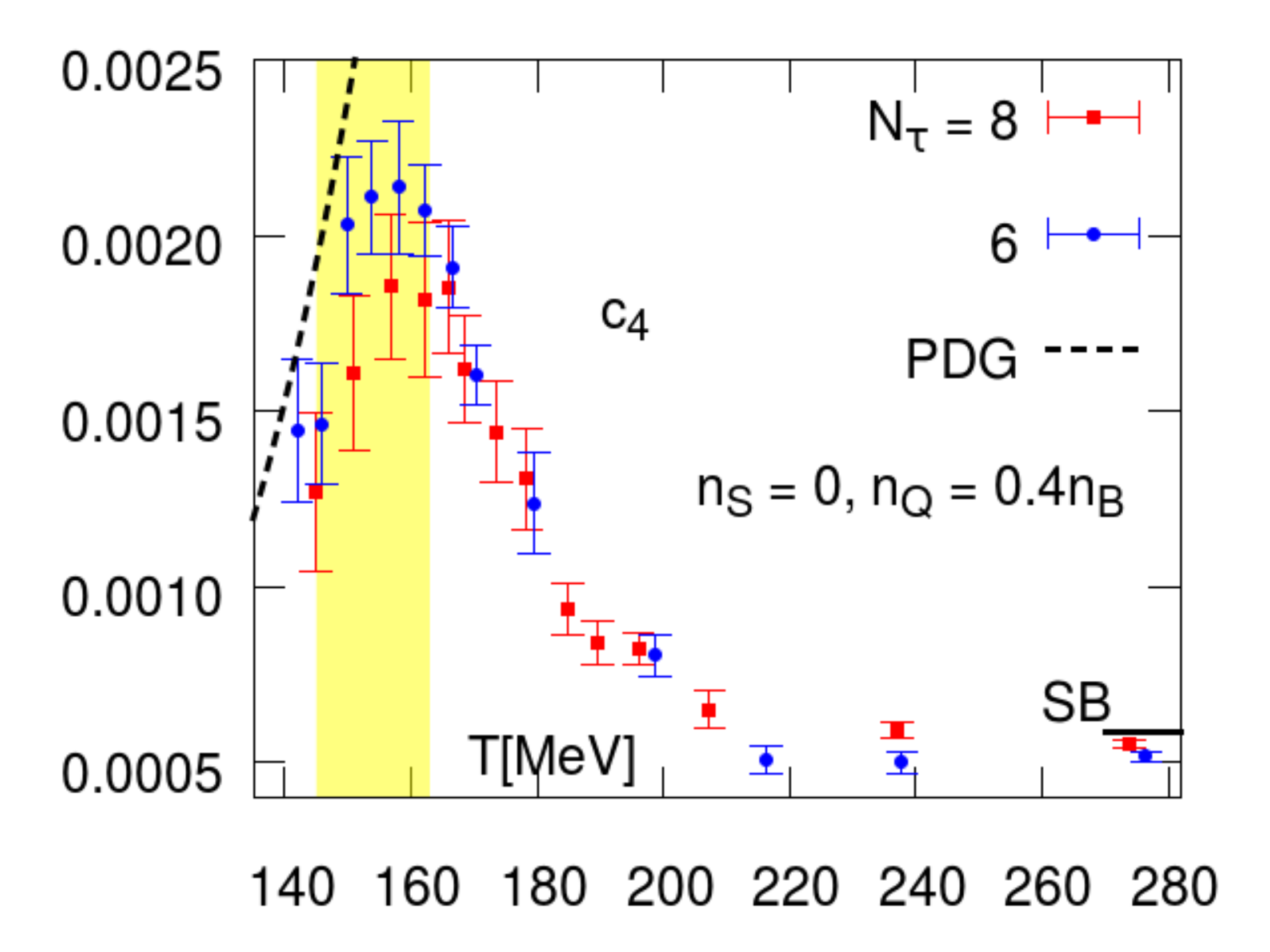}
\includegraphics[width=2.9in]{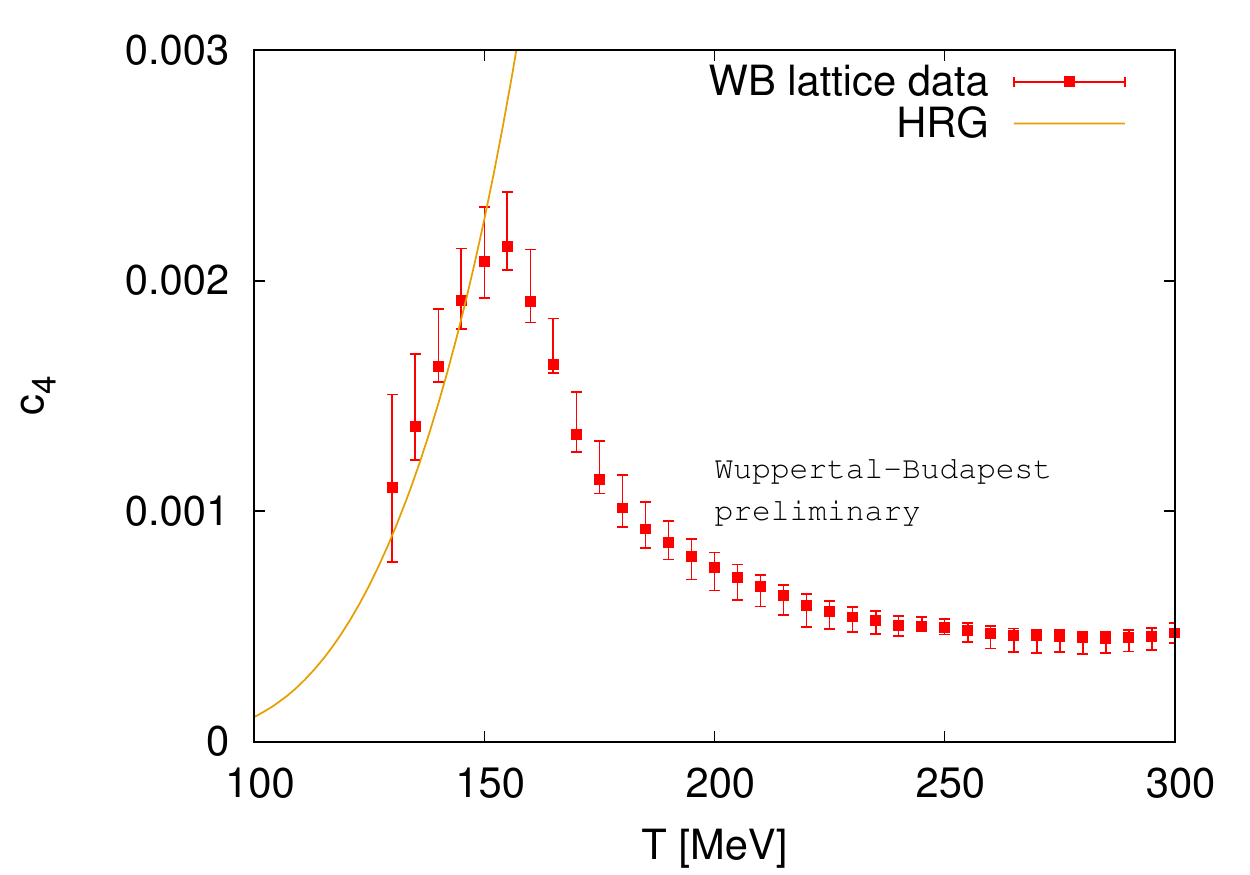}
\end{center}
\caption{\label{fig:c4}
Preliminary $\mu_B^4$ order Taylor coefficient of the QCD pressure.
Left: BNL-Bielefeld group ($N_t=8$) \cite{Hegde:2014wga}, right:
Wuppertal-Budapest group (continuum). The errors are only statistical.
}
\end{figure}

The first continuum result for the leading order Taylor expansion (already
implementing strangeness neutrality) was published in
Ref.~\cite{Borsanyi:2012cr}. The BNL-Bielefeld group has presented
the next orders with $N_t=8$ HISQ fermions \cite{Hegde:2014wga}.
In Fig.~\ref{fig:c4} we show the fourth order coefficient: on the left
panel the BNL-Bielefeld result, on the right panel the preliminary continuum
result of the Wuppertal-Budapest group, based on $N_t=8,10,12$ and 16 lattices
(as of the Quark Matter conference in 2015).

\begin{figure}[ht]
\begin{center}
\includegraphics[width=3in]{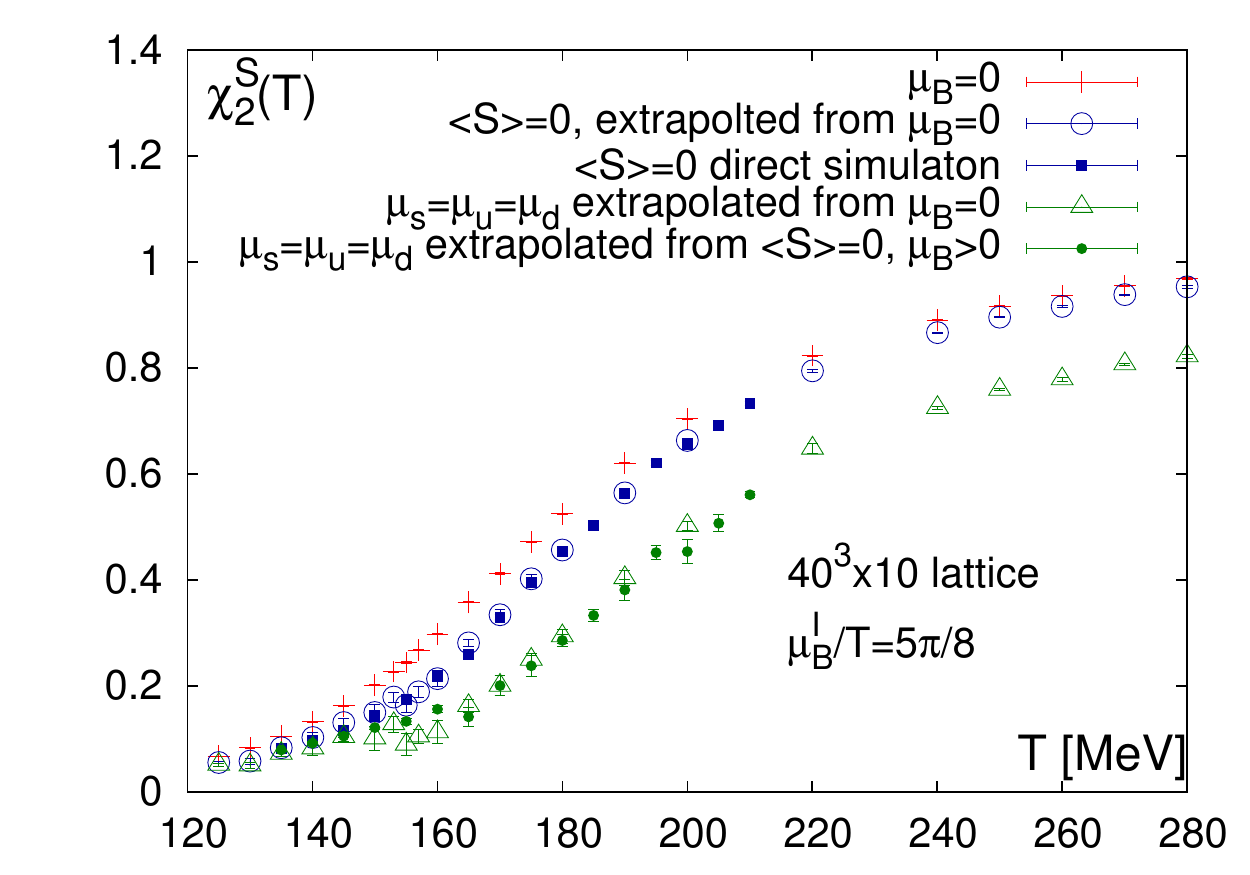}
\end{center}
\caption{\label{fig:c2Sextra}
Strange susceptibility at vanishing (red) and non-vanishing (imaginary)
chemical potentials on a $40^3\times10$ lattice with the 4stout action.
The blue squares are the direct simulation points
at $\mu_B,\mu_S$ pairs where the net strangeness vanishes. These data
are reproduced by a leading Taylor estimate (open circles). If we
keep all chemical potentials equal instead of requiring strangeness
neutrality we arrive at the green dots and triangles. The latter two
refer to an extrapolation from the $\mu_B=0$ or the $\textrm{Im} \mu_B>0$
data, they are in agreement \cite{Bellwied:2015rza}.
}
\end{figure}

Besides the equation of state the $\chi$ coefficients can
also extrapolate the quark number susceptibilities to finite chemical
potentials. The inflection point of the strange susceptibility is one
possible estimator of the deconfinement transition. In Fig.~\ref{fig:c2Sextra}
we show $\chi_2^S(T)$ extrapolated to an imaginary value of the chemical
potential, where direct simulations are also possible. The extrapolation
is to leading ($\chi_B^2$) order. The intrinsic periodicity in the imaginary
part of $\mu_B$ restricts the range of independent simulation points to $0\le
\textrm{Im} \mu_B\le \pi T$. In Fig.~\ref{fig:c2Sextra} the leading order
expansion gives accurate result in more than half of the available interval.

\begin{figure}[ht]
\begin{center}
\includegraphics[height=2in]{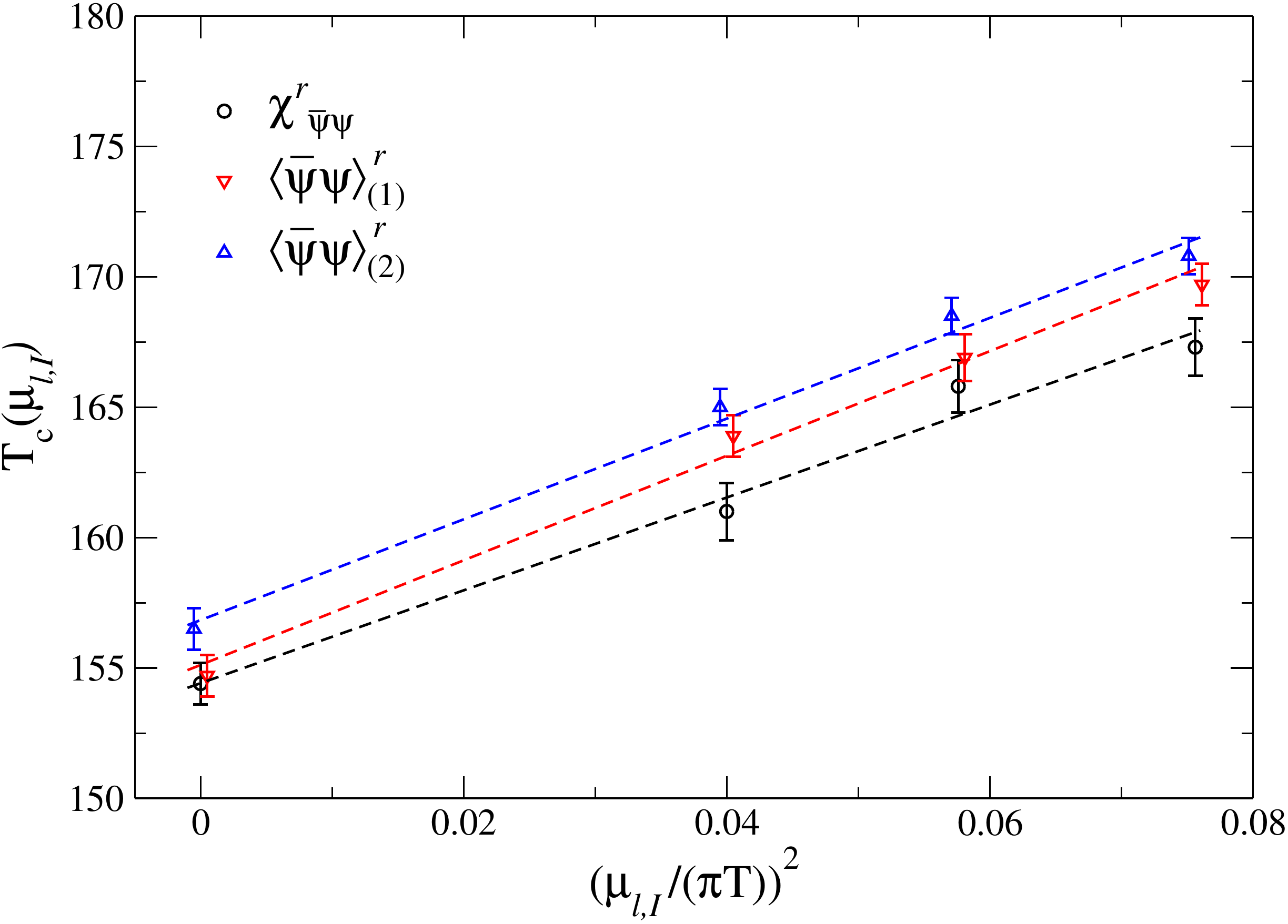}
\includegraphics[height=2in,bb=0 50 545 480]{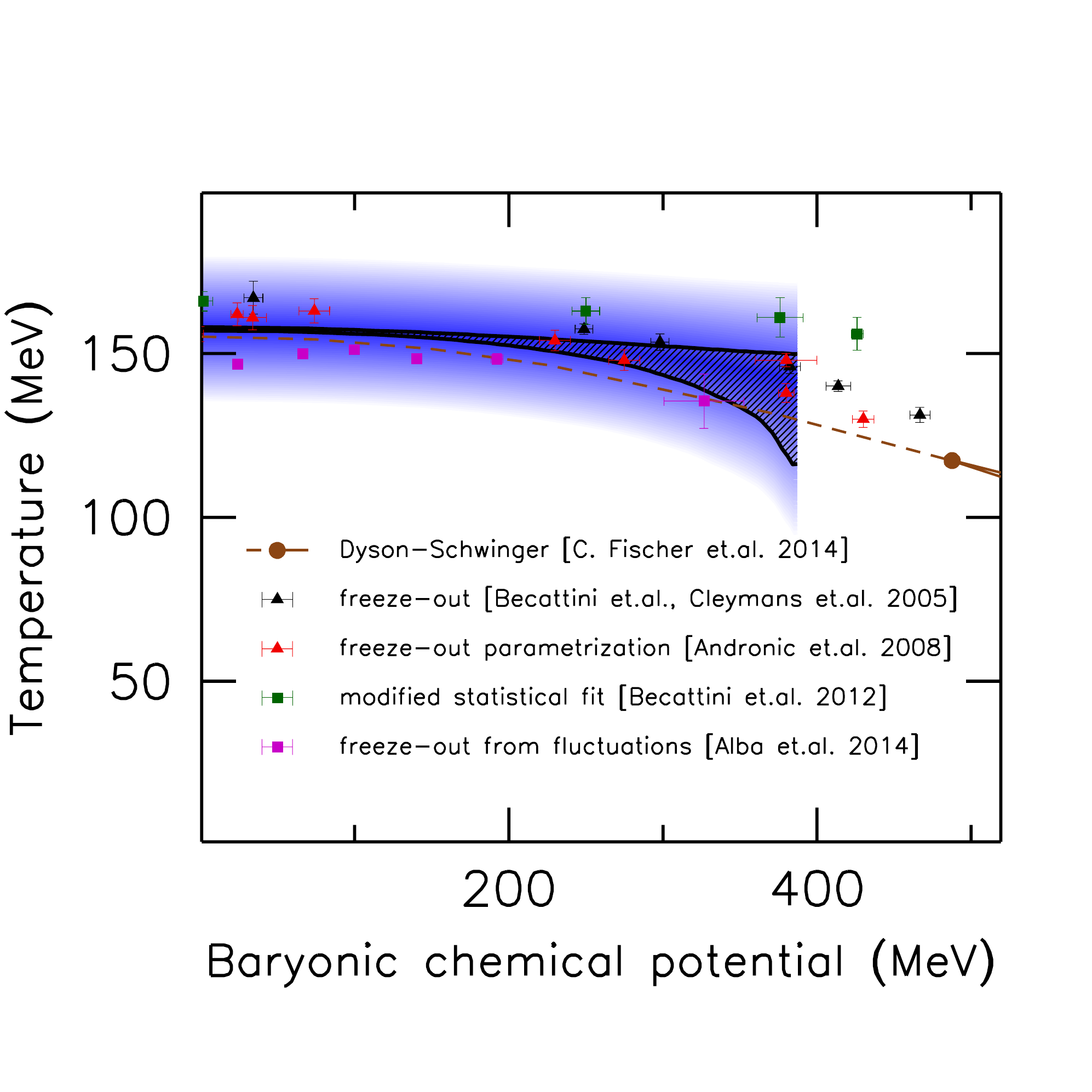}
\end{center}
\caption{\label{fig:pd}
Left: continuum extrapolated 
chiral $T_c$ results at imaginary chemical potentials by the Pisa group
\cite{Bonati:2015bha}. Right: the analytical continuation 
Wuppertal-Budapest group (continuum). The errors are only statistical.
For comparison, we show on the right panel a selection of non-lattice
results: from Dyson-Schwinger equations in Ref. \cite{Fischer:2014ata} 
and from the phenomenological fitting of experimental freeze-out data in
Refs.~\cite{Cleymans:2004pp,Becattini:2005xt,Andronic:2008gu,Becattini:2012xb,Stachel:2013zma,Andronic:2014zha,Alba:2014eba}
}
\end{figure}

The strange susceptibility together with the chiral condensate and
susceptibility was used to calculate the curvature of the transition
line in the QCD phase diagram \cite{Endrodi:2011gv}. The use of imaginary
chemical potentials became a very popular approach, since then the 
$\mu_B$-derivative of the chiral observables do not have to be calculated.
Instead, $T_c$ has to be determined for several imaginary values of the
chemical potentials. Recently three consistent continuum results emerged
\cite{Bonati:2015bha,Bellwied:2015rza,Cea:2015cya}. In Fig.~\ref{fig:pd}
we show the continuum extrapolated $T_c$ results at various imaginary
chemical potentials and the phase diagram after the analytical continuation.
The curvature $\kappa$ of the phase diagram is defined as
\begin{equation}
\frac{T_c(\mu_B)}{ T_c(\mu=0)}=1 - \kappa \left(\frac{\mu_B}{T_c(\mu_B)}\right)^2 + \mathcal{O}(\mu_B^4)\,.
\end{equation}
The Pisa group concluded at $\kappa=0.0135(15)$ (2stout staggered action
up to $N_t=12$) \cite{Bonati:2015bha} the Wuppertal-Budapest group published
$\kappa=0.0149(21)$ (4stout staggered action up to $N_t=16$)
\cite{Bellwied:2015rza}.

%]]] sec:mub

\section{Fluctuations, where theory meets experiment\label{sec:exp}} %[[[

Perhaps the most beautiful aspect of fluctuations of conserved charges is their
availability from heavy ion experiments. Fluctuations are characteristic
to the temperature, chemical potential(s) and volume of a grand canonical
ensemble. Using a somewhat simplified picture, the plasma that was created
at a high energy density equilibrates locally and follows a hydrodynamical
evolution, simultaneously cooling down into the transition range.
Although the total baryon number and electric charge are conserved a 
subsystem can be described by a grand canonical ensemble, though it
is important to consider the finite size of the subvolume \cite{Bzdak:2012an}.
The net abundance of conserved charges in a subsystem is counted by using
rapidity cuts in experiment. The efficiency of the detector is corrected for
and spallation protons are excluded by appropriate cuts in $p_T$, though
such cuts also introduce systematic errors \cite{Karsch:2015zna}.
At RHIC STAR has published results on the first four moments
of the net proton \cite{Adamczyk:2013dal} as well as
net-electric charge \cite{Adamczyk:2014fia} event-by-event fluctuations.
Both STAR and PHENIX have presented further preliminary results at the Quark
Matter (2015) conference. These experiments are part of the beam energy
scan program \cite{Aggarwal:2010cw}, which spans the energy range between 200
and 7.7 GeV center-of-mass beam energies, with a new run $\sqrt{s_{NN}}=14.5$
added this year.

The direct comparison between fluctuations on the lattice and in experiment
have known caveats. First, one assumes the validity of the grand
canonical description in equilibrium. The size ($V$) of the detected subsystem
is an unknown, the measured 
mean ($M\sim V\chi_1$),
variance ($\sigma^2\sim V\chi_2$)
skewness ($S\sim V^{-1/2} \chi_3/\chi_{2}^{3/2}$)
and 
kurtosis ($\kappa\sim V^{-1}\chi_4/\chi_{2}^{2}$)
all carry a power of $V$ as a prefactor. Forming ratios 
\begin{eqnarray}
S\sigma=\chi_3/\chi_{2}
\quad&;&\quad
\kappa\sigma^2=\chi_4/\chi_{2}\nonumber\\
M/\sigma^2=\chi_1/\chi_2
\quad&;&\quad
S\sigma^3/M=\chi_3/\chi_1\,
\label{moments}
\end{eqnarray}
the volume can be cancelled, though a residual $V$ dependence may persist as a
sensitivity of these ratios to the rapidity cut. A further source
of systematics is that $V$ is not constant, its event-by-event
fluctuation mixes into the measured ratios \cite{Skokov:2012ds,Alba:2015iva}.

Nevertheless, we can assume for now, that the fluctuations in experiment
are described by a grand canonical ensemble, that corresponds to the 
last hyper-surface of inelastic scatterings, the chemical freeze-out.
After this the conserved charges are indeed conserved also in the sub-system,
that is finally detected, and its event-by-event fluctuations can
be matched to the QCD prediction. This matching means that a pair
of equations are solved, e.g. \cite{Bazavov:2012jq} 
\begin{eqnarray}
\left.M/\sigma^2\right|_{\mathrm{experiment}} & =& 
\left.\chi_1(T,\mu_B)/\chi_2(T,\mu_B) \right|_{\mathrm{lattice}}\,,\label{eq:mean}\\
\left.S\sigma^3/M\right|_{\mathrm{experiment}} & =& 
\left.\chi_3(T,\mu_B)/\chi_1(T,\mu_B)\right|_{\mathrm{lattice}} \,.\label{eq:skewness}
\end{eqnarray}
This was applied to the electric and baryon charges
\cite{Bazavov:2012jq,Borsanyi:2013hza,Borsanyi:2014ewa}.
In experiment proton fluctuations are measured instead of full
baryon fluctuations, this introduces further systematics 
\cite{Nahrgang:2014fza}. 

In Eqs.~(\ref{eq:mean}-\ref{eq:skewness}) lattice data are needed at finite
chemical potential. These we can have by Taylor-expanding the fluctuations
and use only the highest collision energies (corresponding to small $\mu_B$).
The extrapolation is in all chemical potentials such that
Eqs.~(\ref{eq:neutrality}) are always fulfilled.
At the level of our current precision the small curvature ($\kappa$)
means that the transition temperature is constant within two MeV for
central collisions with $\sqrt{s_{NN}}\gtrsim 27$~GeV (see section
\ref{sec:mub}), the chemical freeze-out temperature is expected to behave
likewise.

\begin{figure}[ht]
\begin{center}
\includegraphics[width=2.9in]{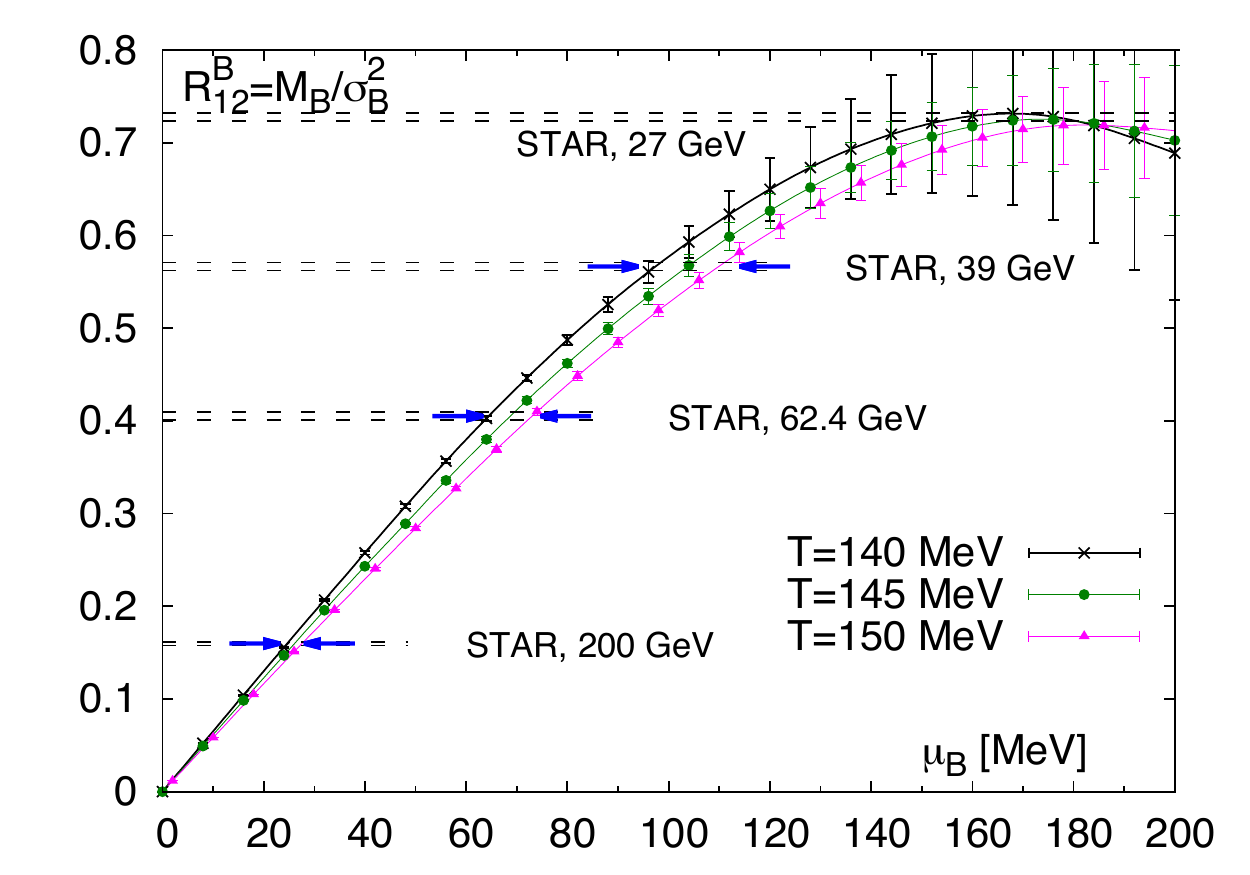}
\includegraphics[width=2.9in]{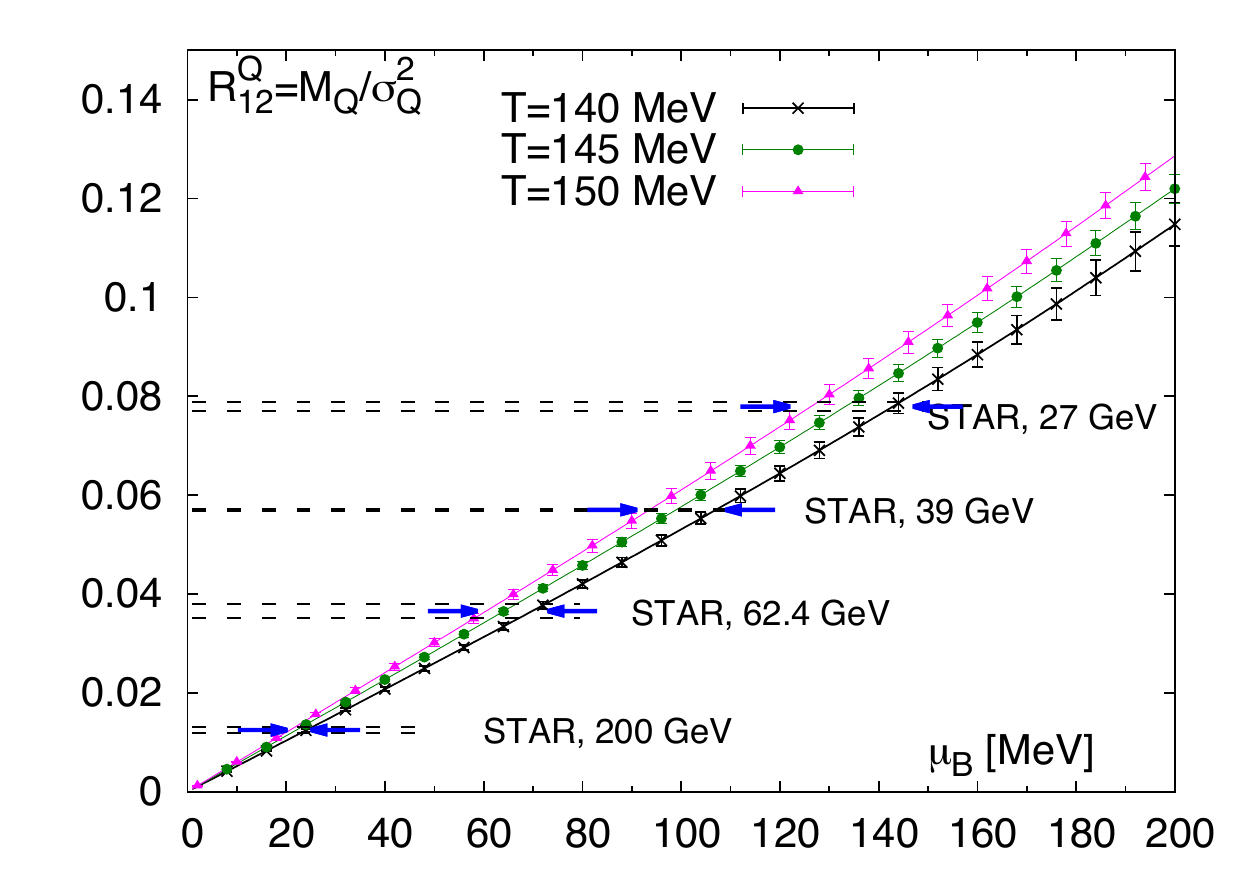}
\end{center}
\caption{\label{fig:mean}
$\chi^B_1/\chi^B_2$ (left) and $\chi^Q_1/\chi^Q_2$ (right) as a function of the
chemical potential using $\mu_B^3$ order Taylor expansion. The STAR
data are from Refs.~\cite{Adamczyk:2013dal,Adamczyk:2014fia} using the
two most central bins (0-10\%) \cite{Borsanyi:2014ewa}.
}
\end{figure}

\begin{figure}[ht]
\begin{center}
\includegraphics[width=2.9in]{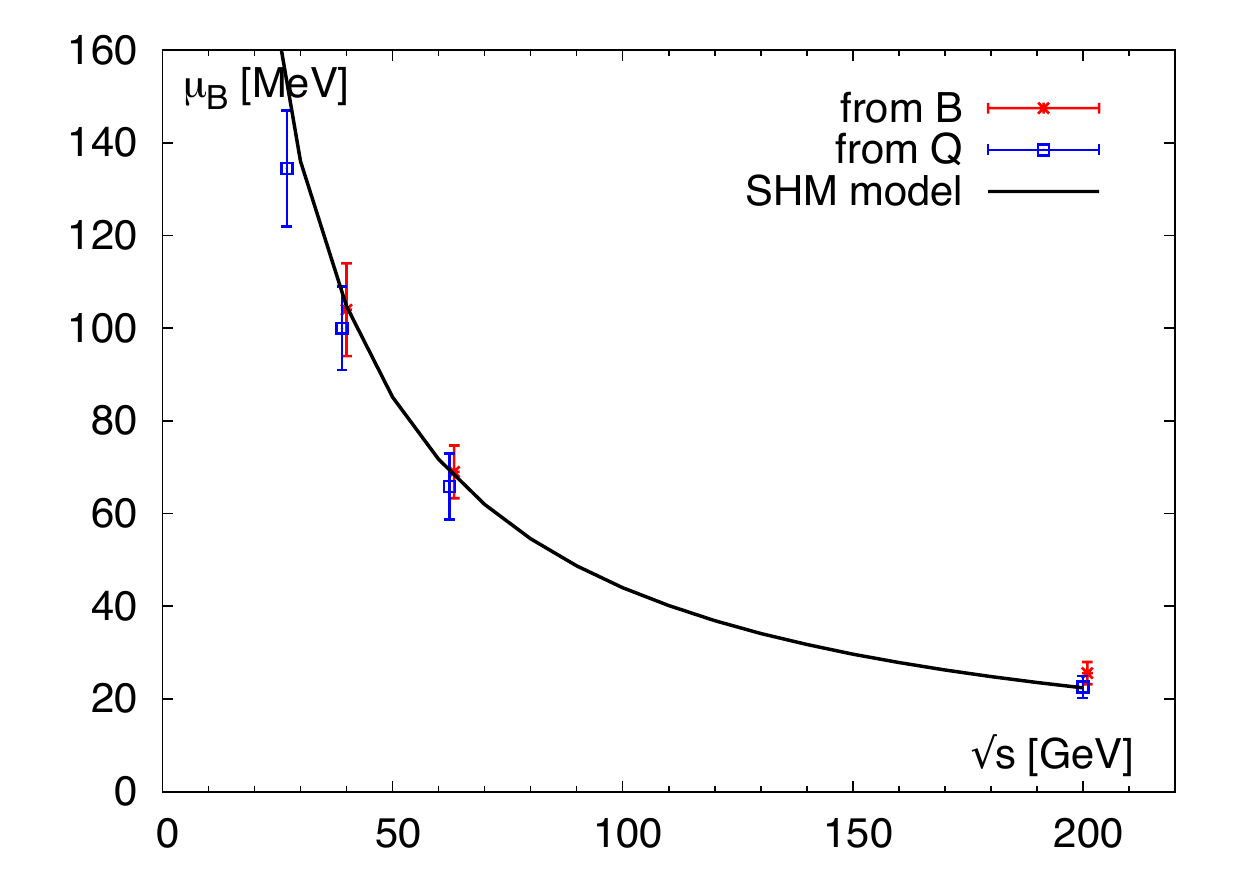}
\includegraphics[width=2.9in]{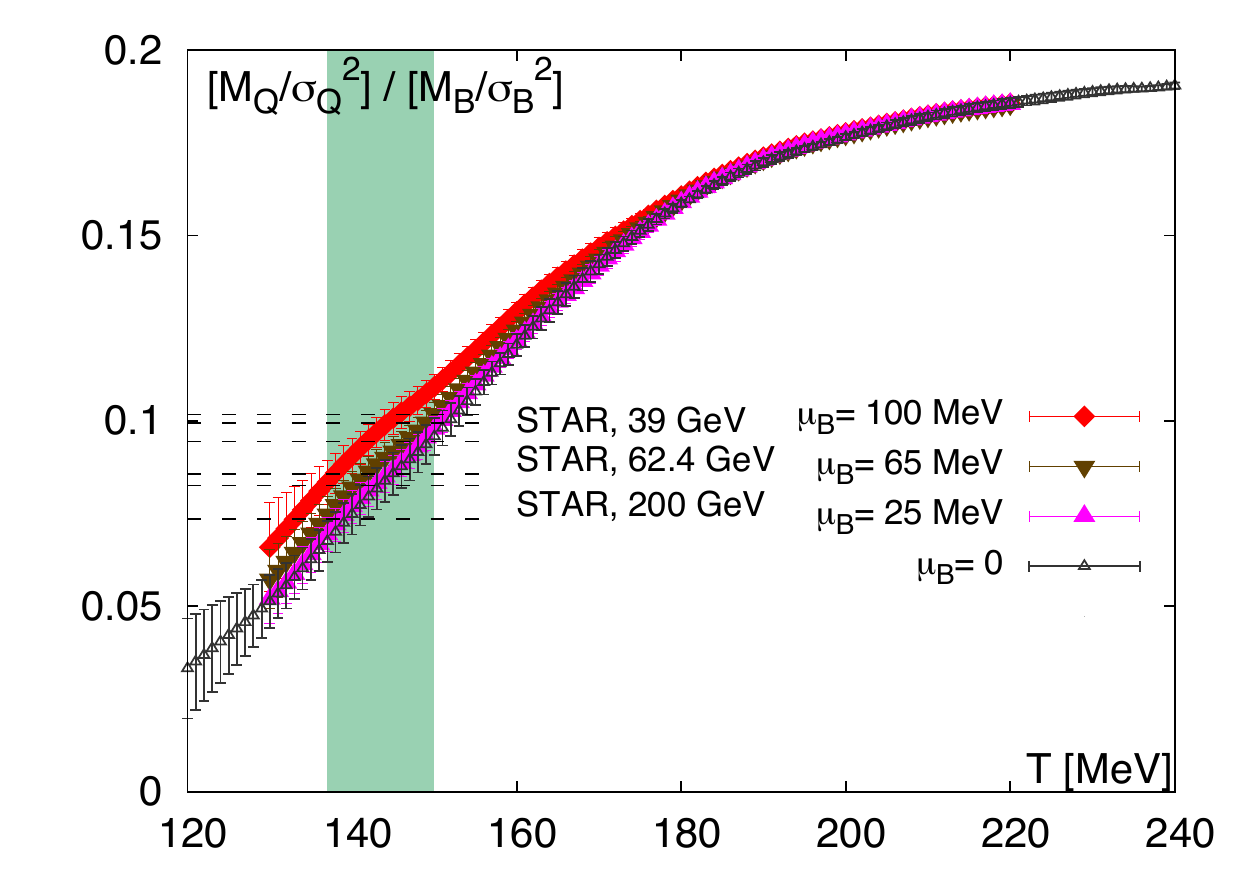}
\end{center}
\caption{\label{fig:freezeout}
Left: comparison of the chemical potentials coming from the freeze-out
analysis to the statistical hadronization model \cite{Andronic:2005yp}.
There is consistency between the baryon and charge-based observables.
If we now assume that $M/\sigma$ correspond to the same ($T,\mu_B$) 
parameters then a thermometer can be constructed from this ratio, which,
in turn is compatible with experimental data only if freeze-out occured
in the transition region, marked with the green band \cite{Borsanyi:2014ewa}.
}
\end{figure}

In Fig.~\ref{fig:mean} we present the $M/\sigma$ ratio for the baryon number
(left) and the electric charge (right). The experimental
data on $M/\sigma^2$ selects a possible range of chemical potentials where
the freeze-out occurred. The shown temperatures span the range preferred
by the skewness data \cite{Borsanyi:2014ewa}. The chemical potentials from the
matching of both data sets to lattice are shown in Fig.~\ref{fig:freezeout}.  
Along with the $\mu_B$ from the Wuppetral-Budapest analysis we show the
result of the Statistical Hadronization Model
\cite{Andronic:2005yp,Andronic:2008gu}. The latter method is the standard
procedure for phenomenological determination of the freeze-out curve, it
compares particle yield ratios to experiment on the basis of the Hadron
Resonance Gas model. Its result is shown with triangles in Fig.~\ref{fig:pd}.
For a similar HRG-based fluctuation analysis see Ref.~\cite{Alba:2014eba}.

A remarkable feature in Fig.~\ref{fig:freezeout} is the agreement between
the two fluctuation data sets and also between the SHM and the
fluctuation approach. This shows the robustness of fluctuations despite
the already discussed systematic effects. We note that SHM gives
higher temperatures then the fluctuation method (either with HRG or with
lattice), this can also be seen in Fig.~\ref{fig:pd}. Notice that the
freeze-out temperatures obtained in the SHM model analysis are in a range where
lattice and HRG model are no longer in agreement.  The non-monotonicity in
Ref.~\cite{Alba:2014eba} may hint at systematic effects that have to be
understood in the future. A recent contribution in this direction calculates
the curvature of the freeze-out line based on the comparison of lattice to
latest STAR data \cite{Bazavov:2015zja}. Though its result is still consistent
with the curvature of the chiral transition line, data prefer a negative value.
As lattice data become increasingly accurate the subtle phenomenological
details of the matching between theory and experiment must be better studied.

%]]]  sec:exp

\section*{Acknowledgement} %[[[

This summary is based on my work in collaboration with R. Bellwied, Z. Fodor,
J. G\"unther, S. D. Katz, S. Krieg, A. P\'asztor,  C. Ratti and K. K. Szab\'o.
The numerical simulations were performed on the QPACE machine, the GPU cluster
at the Wuppertal University, and on JUQUEEN (the Blue Gene/Q system of the
Forschungszentrum Juelich) and on MIRA at the Argonne Leadership Computing
Facility through the Innovative and Novel Computational Impact on Theory and
Experiment (INCITE) program of the U.S.  Department of Energy (DOE).
%]]] Acknowledgement

% Footer [[[
\bibliographystyle{h-physrev.bst}
\bibliography{thermo}{}
%\begin{thebibliography}{99}
%\bibitem{...}
%....
%
%\end{thebibliography}

\end{document}